\documentclass[fleqn,usenatbib,useAMS]{mnras}

\usepackage{newtxtext,newtxmath}

\usepackage[T1]{fontenc}

\DeclareRobustCommand{\VAN}[3]{#2}
\let\VANthebibliography\thebibliography
\def\thebibliography{\DeclareRobustCommand{\VAN}[3]{##3}\VANthebibliography}

\usepackage{graphicx}	
\usepackage{amsmath}	
\usepackage{bm}		
\usepackage[flushleft]{threeparttable}
\usepackage{subcaption}
\usepackage{color}

\newcommand{\blue}[1]{{\color{blue}{}#1}}

\newcommand{\kms}{\,km$\,$s$^{-1}$}
\newcommand{\firb}{f_\mathrm{IRB}}

\title[GASTON: Galactic Star Formation with NIKA2]{GASTON: Galactic Star Formation with NIKA2. Evidence for the mass growth of star-forming clumps}

\author[A. J. Rigby et al.]{
A. J. Rigby,$^{1}$\thanks{E-mail: rigbya@cardiff.ac.uk}
N.~Peretto,$^{1}$\thanks{E-mail: peretton@cardiff.ac.uk}
R.~Adam,$^{2}$
P.~Ade,$^{1}$
M.~Anderson,$^{1}$
P.~Andr\'e,$^{3}$
A.~Andrianasolo,$^{4}$
H.~Aussel,$^{3}$
\newauthor
A.~Bacmann,$^{4}$
A.~Beelen,$^{5}$
A.~Beno\^it,$^{6}$
S.~Berta,$^{7}$
O.~Bourrion,$^{8}$
A.~Bracco,$^{9}$
M.~Calvo,$^{6}$
A.~Catalano,$^{8}$
\newauthor
M.~De Petris,$^{10}$
F.-X.~D\'esert,$^{4}$
S.~Doyle,$^{1}$
E.~F.~C.~Driessen,$^{7}$
P.~Garc\'ia,$^{11,12}$
A.~Gomez,$^{13}$
J.~Goupy,$^{6}$
\newauthor
F.~K\'eruzor\'e,$^{8}$
C.~Kramer,$^{7, 14}$
B.~Ladjelate,$^{14}$
G.~Lagache,$^{15}$
S.~Leclercq,$^{7}$
J.-F.~Lestrade,$^{16}$
\newauthor
J.~F.~Mac\'ias-P\'erez,$^{8}$
P.~Mauskopf,$^{1,17}$
F.~Mayet,$^{8}$
A.~Monfardini,$^{6}$
L.~Perotto,$^{8}$
G.~Pisano,$^{1}$
N.~Ponthieu,$^{4}$
\newauthor
V.~Rev\'eret,$^{3}$
I.~Ristorcelli,$^{18}$
A.~Ritacco$^{19,5}$
C.~Romero,$^{20}$
H.~Roussel,$^{21}$
F.~Ruppin,$^{22}$
K.~Schuster,$^{7}$
S.~Shu,$^{7}$
\newauthor
A.~Sievers,$^{14}$
C.~Tucker,$^{1}$
and E.~J.~Watkins$^{23,1}$
\\
\\
Affiliations are listed at the end of the paper
}

\date{Accepted 2021 January 20. Received 2021 January 20; in original form 2020 July 21}

\pubyear{2020}

\begin{document}
\label{firstpage}
\pagerange{\pageref{firstpage}--\pageref{lastpage}}
\maketitle

\begin{abstract}
Determining the mechanism by which high-mass stars are formed is essential for our understanding of the energy budget and chemical evolution of galaxies. By using the New IRAM KIDs Array 2 (NIKA2) camera on the Institut de Radio Astronomie Millim\'{e}trique (IRAM) 30-m telescope, we have conducted high-sensitivity and large-scale mapping of a fraction of the Galactic plane in order to search for signatures of the transition between the high- and low-mass star-forming modes. Here, we present the first results from the Galactic Star Formation with NIKA2 (GASTON) project, a Large Programme at the IRAM 30-m telescope which is mapping $\approx 2$\,deg$^2$ of the inner Galactic plane (GP), centred on $\ell = 23\fdg9, b=0\fdg05$, as well as targets in Taurus and Ophiuchus in 1.15 and 2.00 mm continuum wavebands. In this paper we present the first of the GASTON GP data taken, and present initial science results. We conduct an extraction of structures from the 1.15\,mm maps using a dendrogram analysis and, by comparison to the compact source catalogues from \textit{Herschel} survey data, we identify a population of 321 previously-undetected clumps. Approximately 80 per cent of these new clumps are 70~\micron-quiet, and may be considered as starless candidates. We find that this new population of clumps are less massive and cooler, on average, than clumps that have already been identified. Further, by classifying the full sample of clumps based upon their infrared-bright fraction -- an indicator of evolutionary stage -- we find evidence for clump mass growth, supporting models of clump-fed high-mass star formation.
\end{abstract}




\section{Introduction}

The stellar initial mass function (IMF) is an essential ingredient in cosmological simulations of galaxy formation and evolution, and its origin remains one of the most fundamental and important questions in the field of astronomy. In the post-\textit{Herschel} era, a link between filamentary structures in the interstellar medium (ISM) and star-formation has been firmly established \citep{Molinari2010, Andre2010, Andre2014}. At the low-mass end of the IMF, most stars with masses in the range $m_{*} \sim 0.1$ to a few $M_{\sun}$ are found to form within dense filaments \citep{Konyves2015, Konyves2020, Marsh2016}. If the mass-per-unit-length exceeds a critical value set by the local sound speed, gravitational instabilities can develop, leading to the fragmentation of the filament, and the formation of stellar systems from this mass reservoir \citep{Inutsuka1997, Clarke2017}. This \textit{core-fed} scheme of fragmentation of trans- or supercritical virialized filaments may explain the origin of the IMF from $\sim 0.1$ to $\sim 5$~$M_{\sun}$ \citep{Lee2017, Andre2019}. However, outside this range, the proposed formation mechanisms for brown dwarfs \citep{Whitworth2007, Luhman2012}, and the formation of high-mass ( $m_{*} \gtrsim 8$\,$M_{\sun}$) stars \citep{Tan2014, Motte2018a}, remain controversial.

It is well established that the Jeans-like fragmentation of molecular clouds is unable to produce dense cores which are sufficiently massive to be the progenitors of high-mass stars \citep{Bontemps2010, Sanhueza2017}, and so the formation pathway must incorporate additional mechanisms. There are broadly two families of models for high-mass star formation that remain actively debated within the field: ones in which the formation of high-mass stars resembles a scaled-up version of the low-mass star formation models \citep[e.g.][]{McKee2003}, where the high-mass protostellar object accretes material from a compact ($\le0.1$\,pc) fixed mass reservoir -- i.e. a core supported by turbulence -- and ones in which protostars grow in mass as a result of the large-scale ($\ge 1$\,pc) and hierarchical gravitational collapse of its parent molecular clump \citep[e.g.][]{Bonnell2006, Vazquez-Semadeni2009, Vazquez-Semadeni2019}. These two families of high-mass star formation models may be  classified as \textit{core-fed} and \textit{clump-fed} scenarios, respectively \citep{Wang2010}.

 A growing weight of observational evidence supports a picture in which large-scale gravitational collapse, resulting in large infall rates, plays a key role in the formation of high-mass stars. Systems with the so-called `hub-filament' configuration \citep{Myers2009} are also routinely found in high-spatial resolution observations of high-mass clumps, with stellar protoclusters \citep{Schneider2012, Liu2016}, or the most massive dense cores found at the filament intersections \citep[e.g.][]{Peretto2013, Peretto2014}. The filaments within these systems have also been seen to exhibit longitudinal velocity gradients that could indicate large-scale gas flows \citep{Kirk2013, Peretto2014, Williams2018, Lu2018, Chen2019} that supply additional kinetic support at the hub centre, whilst funnelling sufficient mass to the high-mass cores on short timescales. However, it is neither clear whether longitudinal filamentary accretion and global collapse in high-mass star-forming hub-filament systems is universal, nor at which masses the transition from the solar-type star forming mode to the high-mass mode occurs.
 
 While the execution of such high-spatial resolution studies across the entire Galaxy remain unfeasible, large and unbiased samples of dense clumps now exist from single dish Galactic Plane surveys \citep{Schuller2009, Aguirre2011, Moore2015, Molinari2016}. The Apex Telescope Large-Area Survey of the Galaxy \citep[ATLASGAL;][]{Schuller2009} at 870~\micron\ identified all of the high-mass clumps ($M > 1000~M_{\sun}$) within the Galaxy \citep{Urquhart2018}, and \citet{Jackson2019} showed that the majority of high-mass clumps are undergoing gravitational collapse. However, limitations in sensitivity or angular resolution prevent these surveys from being able to detect the precursors of intermediate-mass star-forming clumps, where any transition between the core-fed and clump-fed regimes is expected to occur. Studies of clump morphology with higher-angular resolution single dish facilities may allow progress on this front, \citep[e.g.][]{Rigby2018}, but large samples must still be acquired.
 
 In this paper we present the first results from GASTON, a new 200-hour Large Programme (PI: Peretto) being undertaken at the IRAM 30-m telescope using the NIKA2 camera \citep{Bourrion2016, Calvo2016, Adam2018, Perotto2020}. The GASTON project is comprised of three parts aimed at investigating the enigmatic origin of the IMF from different angles: i) a $\sim 2$~deg$^2$ survey of a section of the inner Galactic plane searching for the transition between the solar-type and high-mass star forming scenarios; ii) a search for pre-brown dwarf cores, iii) a study of dust properties in nearby resolved pre-stellar cores. NIKA2 is a state-of-the-art focal plane array at the Institut de Radio Astronomie Millim\'{e}trique (IRAM) 30-m telescope at Pico Veleta in Spain, which completed its commissioning campaign in April 2017. NIKA2 observes in 260~GHz and 150~GHz (hereafter 1.15~mm and 2.00~mm) continuum wavebands simultaneously by means of a dichroic mirror, with angular resolutions of 11.1 and 17.6 arcsec, respectively. There are two 1.15~mm arrays, each made up of 1140 kinetic inductance detectors (KIDs), and there are 616 detectors on the 2.00~mm array, filling the 6\farcm5-diameter instantaneous field of view (FoV) in each case. The noise-equivalent flux densities (NEFDs) are approximately 30\,mJy\,s$^{1/2}$ at 1.15\,mm and 9\,mJy\,s$^{1/2}$ at 2.00\,mm which, when combined with the large instantaneous FoV, result in mapping speeds that are an order of magnitude larger than the previous generation of instrumentation in operation at the IRAM 30-m telescope.
 
 In this paper, we describe the observations and present the first results from the GASTON GP project, in which we explore the potential for the NIKA2 observations to identify signatures of the core-fed or clump-fed scenarios of high-mass star formation. The paper is structured in the following way. In Sect. \ref{sec:obs}, we describe the observations and data reduction, while in  Sect. \ref{sec:analysis} we describe our analysis of the 1.15~mm data and present the results, in Sect. \ref{sec:results}. We summarise our results in Sect. \ref{sec:conclusions}

\section{Observations} \label{sec:obs}

\subsection{Observing strategy}

\begin{figure*}
    \centering
    \includegraphics[width=\textwidth]{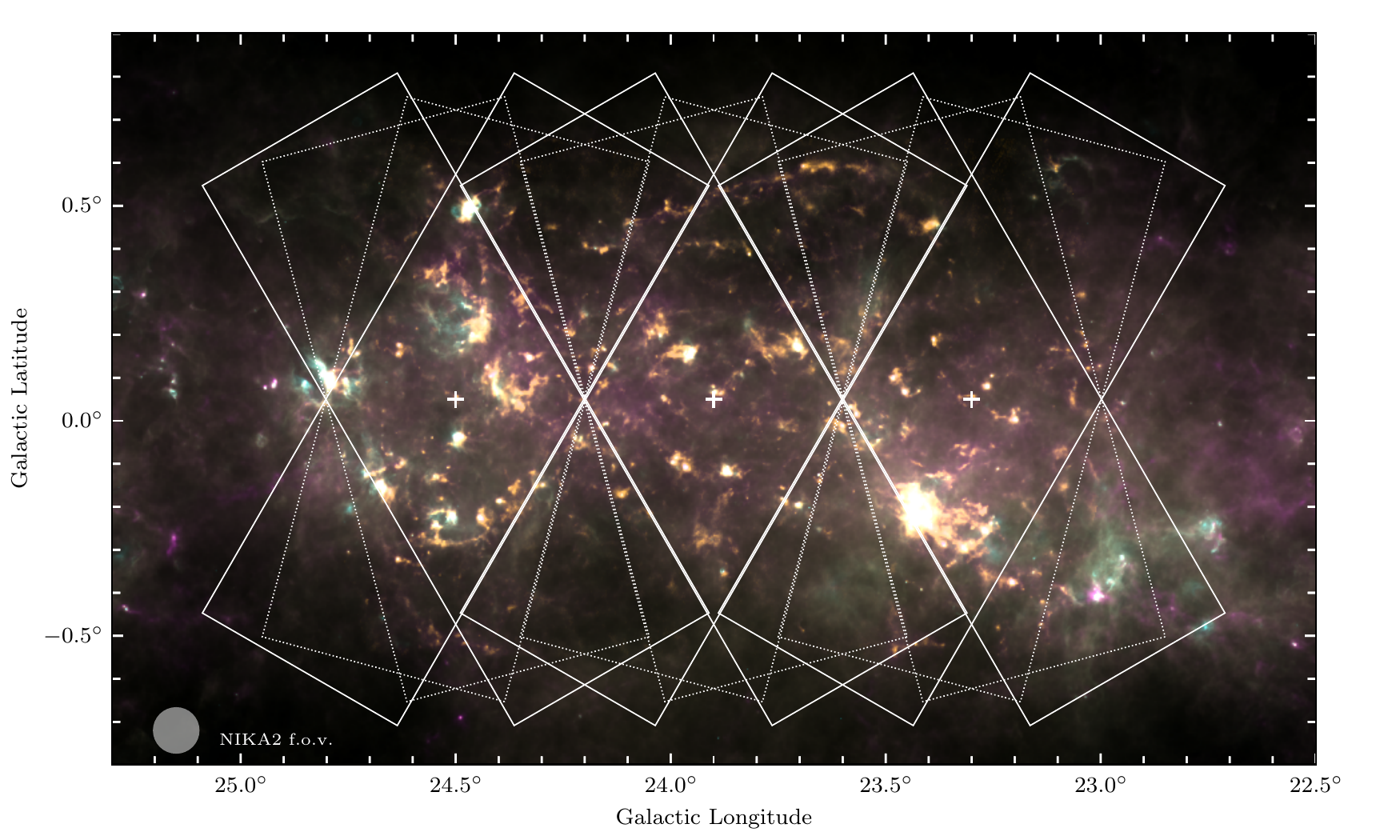}
    \vspace*{-4mm}
    \caption{Scanning strategy for the GASTON GP field. Rectangles show the area covered by the central pixel of the arrays in the various scans centred on the white crosses. Dotted white rectangles show the initial strategy, and solid white rectangles show the final adopted strategy. The background image is a four-colour composite\protect\footnotemark consisting of \textit{Herschel} Hi-GAL \citep{Molinari2016} 70\,\micron, 160\,\micron, and 350\,\micron\ images in cyan, yellow, and magenta, and NIKA2 1.15\,mm signal-to-noise ratio (to exclude noisy edges) in orange (this study). The circle in the lower left corner shows the extent of the NIKA2 FoV.}
    \label{fig:strategy}
\end{figure*}

\footnotetext{Image generated using the {\sc python} package {\sc multicolorfits}: \url{https://github.com/pjcigan/multicolorfits}}

The GASTON GP field, centred on $\ell = 23\fdg9$, $b = 0\fdg05$, was selected for the project due to its extremely high density of compact dust continuum sources, as identified by the \textit{Herschel} infrared Galactic Plane Survey \citep[Hi-GAL;][]{Molinari2016}, as well as a large of number of infrared-dark clouds \citep[IRDCs;][]{Peretto2009}. The GP part of GASTON has been allocated a total of 73.6 hours of telescope time. The approximately 2 deg$^2$ field was divided into three sub-maps, centred on $\ell = 23\fdg3$, $23\fdg9$, and $24\fdg5$, and $b=0\fdg05$, which were observed using an on-the-fly mode, and combined to form the survey area. Pairs of observations with alternating position angles at these three positions are repeated in order to obtain the total integration time, and in such a manner that an approximately uniform sensitivity should be achieved over a large fraction of the map. 

The pairs of rectangular raster-scanned maps, are performed with alternating position angles in an effort to reduce striping in the reduced maps and are scanned across the plane, as illustrated in Fig. \ref{fig:strategy}, to ensure that the start-point of each scan is in a position of low emission. Initially, we adopted a strategy in which scans were alternated by $\pm 15$\degr\ with respect to the normal to the Galactic plane, as used by ATLASGAL \citep{Schuller2009}, but a $\pm 30$\degr\ strategy was finally adopted for most scans, representing a compromise between reducing striping artefacts from data reduction, and maintaining a good mapping efficiency. For the first of the observing runs (N2R12 and N2R14), scans were made with a length of 78 arcmin, with 195 arcsec row spacing (i.e. one half of the 6\farcm5 FoV) between a total of 11 adjacent sub-scans, and with a scanning speed of 60 arcsec per second. Following the second observing run (N2R14), some minor adjustments were made: i) the scanning speed was increased to 70 arcsec per second, in order to assist with the recovery of extended emission; ii) the sub-scan spacing reduced to 180 arcsec to achieve more consistent coverage between sub-scans; iii) the sub-scan length was changed to 87 arcmin to achieve the same latitude coverage as the earlier scans.

The maps presented in this study are the result of a combined 30 hours of integration time, and consist of a total of 108 individual observations taken over five observing pools: N2R12 (2017 October 27--30), N2R14 (2018 January 17--23), N2R15 ( 2018 February 17--18), N2R26 (2019 January 15--22), and N2R28 (2019 February 12--19). The observations were carried out with sky opacities ranging from $\tau_{225 \ \mathrm{GHz}}=0.08$ to 0.44, and a mean value of 0.19, as measured by the IRAM 30-m taumeter. The mean source elevation was 39\fdg9, and the majority ($\gtrsim 80$ per cent) of the observations  were obtained between the hours of 08:00--14:00, indicating that the primary beams are likely to be relatively stable around their nominal values of 11.1 (1.15~mm) and 17.6 (2.00~mm) arcsec \citep[][se their Fig. 12]{Perotto2020}. The remainder were observed between 15:00--19:00, during which time the primary beams tend to degrade slightly to $\sim 12.5$ and 18.0 arcsec, respectively. A detailed description of the operation, calibration and data reduction methods for NIKA2 can be found in \citet{Adam2018} and \citet{Perotto2020}.

\subsection{Data reduction}

The data were reduced using the IDL pipeline developed by the NIKA Core Team, which converts the raw time-series data into calibrated maps. While a detailed description of this pipeline will be presented in \blue{Ponthieu et al. }(\blue{in prep}), we summarise the main steps here.

First, individual KIDs that do not meet the performance criteria, are masked from the timelines, as is also done for samples associated with detected cosmic ray impacts. Next, the pointing information from the telescope control system is used in conjunction with a record of the positional offsets of each KID with respect to a reference position to determine the position of each KID on the sky as a function of time. The KIDs are then inter-calibrated, using coefficients determined from their relative gains, before applying a correction for the instantaneous atmospheric opacity, and finally applying an absolute calibration determined for the specific observing run (see Sect. \ref{sec:calibration}). 

At this stage, a model of the correlated low-frequency noise components, which consist of electronic noise that is coupled to the readout electronics of subsets of KIDs, and atmospheric noise, which is common to all KIDs, is created; this is the so-called `common-mode'. We use the `Most Correlated Pixels' method \citep{Perotto2020}, which works by first discarding any samples in each timeline that, when combined with the telescope pointing information and a source mask, are thought to coincide with an astrophysical source. Next, cross-correlation coefficients between all KIDs are determined and, for each KID, a model of the low-frequency noise is built by co-adding the timelines of that KID and the 15 most correlated KIDs. A linear fit of this common mode is made and subtracted from each timeline, leaving, in principle, the true astrophysical signal.

Since the GASTON GP field is known to contain a large number of bright and diffuse structures with complex morphologies, we adopt an iterative method to define the source mask used in the calculation of the common mode. For the initial iteration, we use a source mask created\footnote{The source mask is created by first rescaling the 500~\micron\ Hi-GAL map to 1.15\,mm, assuming a greybody spectral energy distribution with a dust emissivity spectral index of $\beta = 1.8$ and a dust temperature of 12~K. We then subtract a version of those data smoothed to 6.5 arcmin to approximate the NIKA2 spatial filtering, and convert the resulting map into a binary source mask by identifying those pixels which exceed half of the expected sensitivity of 1.55 mJy beam$^{-1}$.} from a Hi-GAL map of the region at 500~\micron\ \citep{Molinari2016}. On subsequent iterations, the source mask is determined from the astrophysical signal from the previous iteration, masking out all samples where the signal-to-noise ratio (S/N) is greater than 2, and the process is repeated to improve the estimate of the common mode, and thereby the astrophysical signal. We used a total of 55 iterations, at which point, the change in the flux is negligible compared to the map produced by the penultimate iteration.

Finally, the timelines, when combined with the telescope pointing information, are projected (tangential projection) onto a pixel grid with 2.5 arcsec pixels. Inverse-variance noise weighting is used to assign the data samples onto the pixel grid, using a nearest grid point method, and to combine the maps associated with each individual scan into a mosaic of the GP field. 

The correlated noise removal is the origin of the spatial filtering within the reduced maps, which is common to all submillimetre and millimetre-continuum imaging from ground-based telescopes. The level of spatial filtering applied here is typically on the order of the instantaneous FoV which, for NIKA2, is approximately 6\farcm5. We do not calculate the pipeline transfer function that describes the ratio of the power spectrum within the reduced image to an estimate of the absolute source power spectrum over the sky region in this paper. An example of the 2.00~mm NIKA2 transfer function towards a galaxy cluster may be found in \citet[][see their Fig. 3]{Ruppin2018}, who found that around 95 per cent of emission at angular scales between the 2.00 mm beam size and the FoV is recovered, falling off rapidly at larger scales, and we expect a similar level of recovery here.

\subsection{Calibration} \label{sec:calibration}

Uranus and Neptune are used as the primary flux calibrators, and so colour corrections must be applied to our photometric maps to account for the difference between the spectral energy distributions (SEDs) of those planets, and the approximate SEDs of the sources is the NIKA2 imaging. We apply colour correction factors of 1.02 and 0.97 to the 1.15 and 2.00 mm bands, respectively, which were interpolated from Table 12 of \citet{Perotto2020}, assuming an SED of the form $I_\nu \propto (\nu/\nu_0)^{2 + \beta}$, with the mean Galactic plane value for the dust emissivity spectral index of $\beta = 1.8$ \citep{PlanckCollaboration2011}. The total calibration uncertainties are determined from the quadrature sum of the 5 per cent planet model uncertainties, the 5.7 and 3.0 per cent point-source rms calibration uncertainty, the 0.6 and 0.3 per cent systematic calibration uncertainties, and the 3 and 2 per cent uncertainties in the reference beam efficiencies, giving a total of 8 and 6 per cent calibration uncertainty on the 1.15 and 2.00 mm integrated flux densities, respectively. A detailed description of the absolute calibration procedure can be found in \citet{Perotto2020}.

\subsection{Ancillary data} \label{sec:linedata}

We make use of two spectral line data sets in order to determine velocities to continuum sources in Sect. \ref{sec:velocities}. Firstly, we use $^{13}$CO (1--0) data (110.201 GHz) from the FOREST unbiased Galactic plane imaging survey with the Nobeyama 45~m telescope \citep[FUGIN;][]{Umemoto2017}, which have an angular resolution of 21 arcsec, and an rms of $\sigma(T_\mathrm{A}^{*}) = 0.87$~K per 0.65\kms\ velocity channel. The spatial extent of the FUGIN data covers the entire GASTON GP field, and the velocity range of $-100 < v_\mathrm{LSR} < 200$\kms, covers all of the Galactic spiral arms along this sight-line \citep{Dame2001}.

We also make use of data from the pilot study for the Radio Ammonia Mid-Plane Survey \citep[RAMPS;][]{Hogge2018}, a Galactic plane survey of ammonia and water maser emission carried out with the Green Bank Telescope. The L23 and L24 regions of the pilot study have a considerable overlap with the GASTON GP field, spanning $22\fdg5 \gtrsim \ell \lesssim 24\fdg5$, with $|b| \lesssim 0\fdg4$, and we make use of the moment 1 maps of NH$_3$ (1,1) inversion emission at 23.694 GHz. These data have an angular resolution of 32 arcsec, and 0.018\kms\ spectral channels.

Data from Hi-GAL \citep{Molinari2016} at 160 and 250~\micron\ are used in Sect. \ref{sec:properties} for colour temperature determinations. The 160~\micron\ PACS \citep{Poglitsch2010} data have an angular resolution of 12 arcsec, and the 250~\micron\ SPIRE \citep{Griffin2010} data have an angular resolution of 18 arcsec. In both cases the uncertainties are dominated by calibration, which we take as 7 per cent for PACS\footnote{\url{https://www.cosmos.esa.int/web/herschel/pacs-overview}} and 6.5 per cent for SPIRE\footnote{\url{https://www.cosmos.esa.int/web/herschel/spire-overview}}, including a 1 per cent contribution for extended sources due to uncertainty in the SPIRE beam.

We also use 8~\micron\ data from the Galactic Legacy Infrared Mid-Plane Survey Extraordinaire \citep[GLIMPSE;][]{Benjamin2003, Churchwell2009} in Sect. \ref{sec:evolution}. These \textit{Spitzer} data have an angular resolution of 2.0 arcsec.

\section{Analysis} \label{sec:analysis}

\subsection{Photometric maps}

\begin{figure*}
    \centering
    \includegraphics[width=\textwidth]{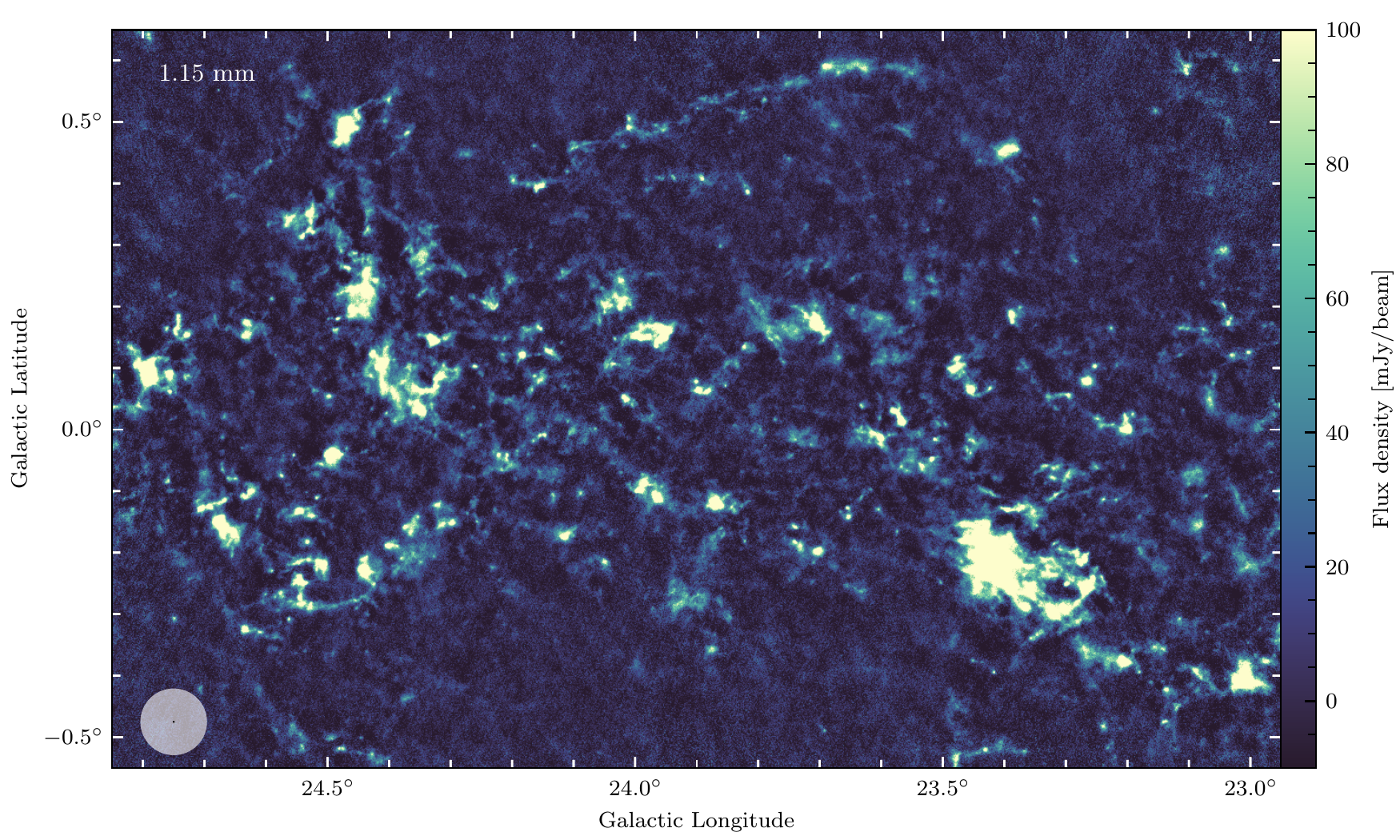}
    \includegraphics[width=\textwidth]{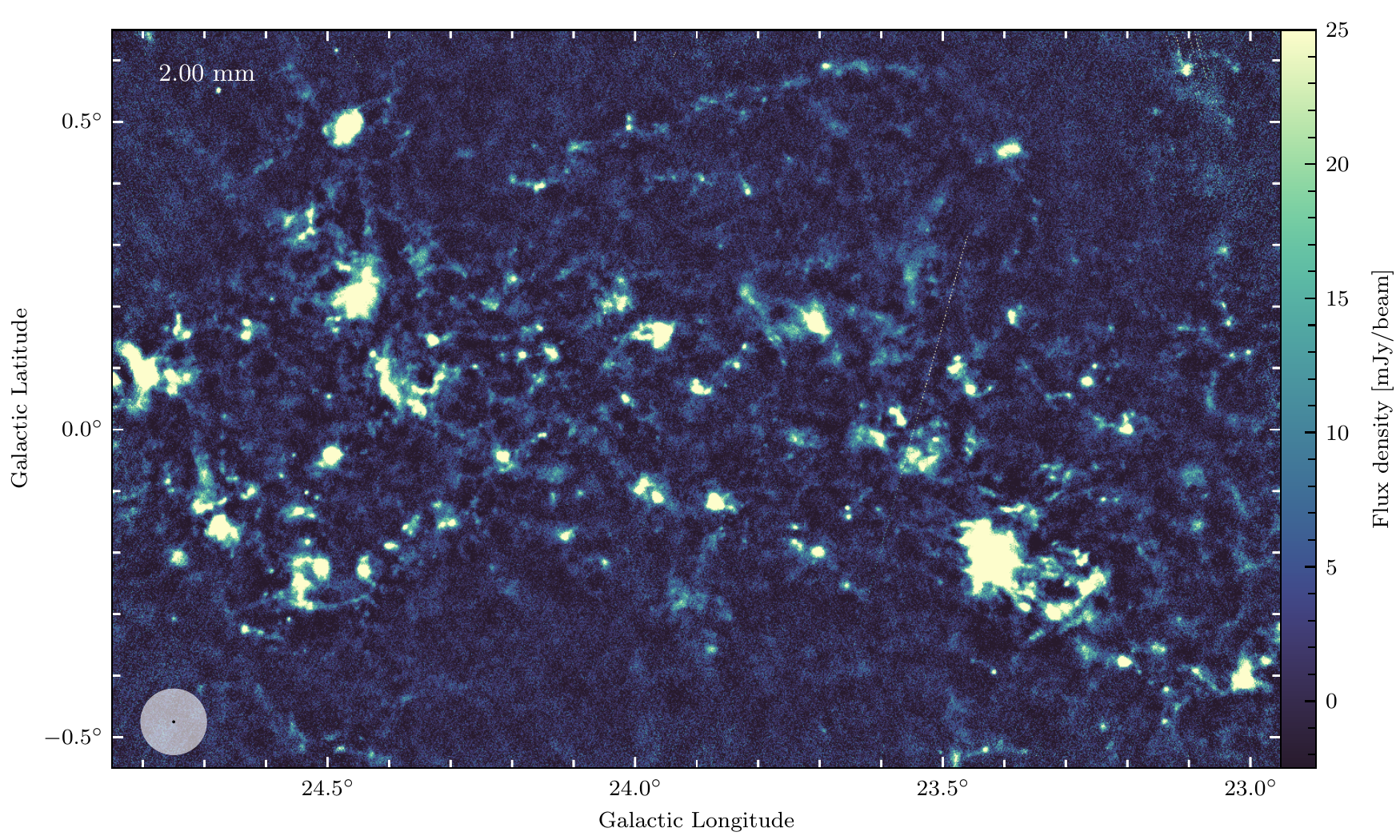}
    \caption{GASTON GP field at 1.15\,mm (top panel) and 2.00\,mm (bottom panel), which have FWHM resolutions of 11.1 and 17.6 arcsec, respectively. The FoV and primary beam sizes are shown in the bottom left corners as opaque white, and black circles, respectively.}
    \label{fig:maps}
\end{figure*}

In Fig. \ref{fig:maps}, we present the NIKA2 photometric maps, after cropping to the largest contiguous rectangular field where the sensitivity is approximately uniform. The total integration time obtained so far is $\sim $30\, hours in 108 scans, corresponding to roughly 40 per cent of what will be acquired upon the completion of GASTON. 

Although we do not use the 2.00\,mm map in the analysis presented in this paper, we present it in Fig. \ref{fig:maps} for the sake of completeness, and to display the data quality. We will make use of these data in future GASTON studies and, in particular, the 1.15\, and 2.00\,mm data will be used in combination to study the dust emissivity spectral index, $\beta$, as in \citet{Rigby2018}.

\subsection{Source extraction} \label{sec:sourceextraction}

A source extraction was carried out on the 1.15\,mm maps using the {\sc python} {\sc astrodendro}\footnote{\url{http://www.dendrograms.org/}} package, which uses a dendrogram \citep{Rosolowsky2008} scheme to segment the emission while maintaining the ability to track its hierarchical structure. The dendrogram separates emission features when they are considered to be significant given their size, and difference in their intensity with respect to the background and to overlapping emission features.

The 1.15\,mm map was first cropped to $22\fdg95 \leq \ell \leq 24\fdg85$ and $-0\fdg55 \leq \ell \leq 0\fdg65$ (as in Fig. \ref{fig:maps}) in order to exclude the noisier edge regions. The map was then smoothed to 13 arcsec resolution in order to increase the S/N. Measuring the sensitivity in such deep maps of the inner Galactic plane is problematic because of source confusion, and so several methods were used to measure the rms noise level. The preferred method was to perform a Gaussian fit to the distribution of pixel values taken from the null maps within a sample of four octagonal sky apertures with radii of $\sim$1.9 arcmin (displayed in Appendix \ref{app:nullmaps}) located across the main survey region. The four sky apertures were carefully placed across the map in order to both sample the range of integration depth in a representative manner, and to avoid regions with residual emission. The null maps are produced by the data reduction pipeline, and are made by alternately adding and subtracting the maps generated by individual scans of the full observation set. In principle, this produces a map free of emission structures whilst preserving the noise characteristics of the map. This method yields the most robust noise estimate in a line-of-sight with emission structures at all positions. The rms determined in this way was 4.18 mJy~13-arcsec-beam$^{-1}$.

Based on the estimated rms noise level, we set the minimum value ({\tt min\_value}) above which emission is considered to be real, and the minimum difference between substructures ({\tt min\_delta}) to 3 times the global rms, while the minimum structure size ({\tt min\_npix}) to be 16 pixels -- equivalent to half of the 13-arcsec beam area. The latter choice was made for the minimum size criterion because pure point sources will be reported as smaller than the beamsize due to the clipping of the wings that extend below the minimum intensity level (as in the `bijection' scheme described in \citealt{Rosolowsky2008}), and so would be filtered from the data if the full beam area was used. This is especially true for low S/N sources.

The application of a single average rms noise level over the entire area results in a small number of spurious detections in the noisier-than-average regions and, conversely, some undetected compact sources are visible in the residuals in the regions where the sensitivity is greater than this average value. To enable us to eliminate spurious sources, we calculate the local S/N of each source in each waveband. As mentioned earlier, the null maps contain some residual positive or negative emission, and are unsuitable for use as noise maps for all sources. We therefore produce a noise map in each waveband, that is calculated from the weight maps produced by the IDL pipeline. The per-pixel weights are proportional to the inverse variance of the flux densities measured at each position on the pixel grid from the contributing observations. The noise at each pixel position, $\sigma_i$, is therefore calculated as $\sigma_i = \sigma_{0} \sqrt{w_0 / w_i}$, where $\sigma_0$ is the rms measured from the pixels within the four sky apertures, $w_0$ is the mean value of the pixel weights within the four sky apertures, and $w_i$ is the per-pixel weight. We apply a `pruning' scheme in the dendrogram construction such that any objects that do not have a local S/N that exceeds 6 -- in keeping with the dendrogram's input parameters -- in at least one of their constituent pixels are removed.

In the dendrogram nomenclature, emission features may described as `trunks', `branches', or `leaves', depending on their position within the hierarchy. The trunks are the lowest-lying structures in the hierarchy, at the minimum detection threshold and thus describe the maximum extent of any isolated emission region, while branches are at a higher contour level within a trunk, and may contain multiple leaves. In the following analysis, we refer to the dendrogram leaves -- those emission features that contain no further discernible substructure -- as `clumps', whereas we use the term `source' to refer to any emission feature extracted using the dendrogram, regardless of its position within the hierarchy. That is to say that a source may contain multiple clumps. The dendrogram extraction found a total of 2446 sources of all kinds, consisting of 488 isolated emission structures (trunks) with 1467 clumps (leaves). We summarise the basic properties -- flux densities and source sizes -- in Appendix \ref{app:catalogue}.

\subsection{Source velocity assignments} \label{sec:velocities}

Determining the systemic velocity, and thereby a kinematic distance, for each extracted source is a necessary step in calculating many of its physical properties. However, due to both the high sensitivity of the GASTON GP maps at 1.15\,mm, and the particularly complicated line-of-sight for the survey region, source confusion presents a significant challenge in their determination. We therefore adopted a procedure with several stages to establish the most reliable possible velocities, which are summarised in this Section, and further details may be found in Appendix \ref{app:velocities}.

\begin{table*}
    \centering
    \caption{Distribution of quality flags for each stage in the velocity determination process. For each flag value, the total number of sources with that value ($N$) and the percentage of the sample made up by those sources (\%) is given.}
    \begin{tabular}{@{\extracolsep{6pt}}clcccccc@{}}
    \hline
    Stage & Description        & \multicolumn{6}{c}{Flag value} \\ \cline{3-8}
    &  & \multicolumn{2}{c}{3} & \multicolumn{2}{c}{2} & \multicolumn{2}{c}{1} \\ \cline{3-4} \cline{5-6} \cline{7-8}
              &   & $N$ & \%   & $N$ & \%   & $N$ & \%  \\
    \hline
    i) & FUGIN $^{13}$CO (1--0)  & 1685 & 68.9 & 298  &  12.2 &  463 & 18.9\\
    ii) & RAMPS NH$_3$ (1,1) & 1916 & 78.3 & 165  &  6.7  &  365 & 14.9\\
    iii) & Dendrogram refinement & 2169 & 88.7 & 75   &  3.1  &  202 & 8.3 \\
    \hline
    \end{tabular}
    \label{tab:vflags}
\end{table*}

To determine a velocity for each GASTON source, first, an integrated $^{13}$CO (1--0) spectrum is extracted from the FUGIN data of all pixels that lie within the source footprint. Many of the spectra exhibit multiple emission features, and so the velocity with the greatest intensity is assigned as the centroid velocity, $v_\mathrm{cen}$, of each source. In cases where the line of sight contains multiple velocity components, this velocity is assumed to trace the highest-column density material which is most likely to be associated with the dust continuum source. A quality flag was assigned to each velocity assignment, with a value of 3 given to the most robust assignments, and 1 to the poorest, and a value of 2 is given for intermediate quality assignments.

The main advantage of this method is its simplicity, but it may obviously provide wrong velocity assignments in some cases, primarily due to the difference in densities traced by the molecular and continuum data. In principle, the dust continuum emission traces the full column density of molecular hydrogen, but the spatial filtering applied by the data reduction pipeline removes much of the emission from diffuse gas, leaving the more compact, and higher column-density gas. Observations of dense gas tracers would provide more suitable velocities for the GASTON GP sources than $^{13}$CO (1--0), and so, as a second stage, we use data from the RAMPS pilot study \citep{Hogge2018} of NH$_3$ (1,1), which has a critical density of $n_\mathrm{crit}\sim 3 \times 10^3$~cm$^{-3}$ to determine dense-gas velocities. We adopt velocities for 517 (21 per cent) of the sources from the first component from the line-fitting procedure, and a further 513 (21 per cent) from the first-moment maps of \citet{Hogge2018}, assigning a quality flag of 3 for all ammonia-derived velocities.

The hierarchical structure of the 1.15\,mm emission is recorded within the dendrogram catalogue, and since each source now has an individual velocity assignment, along with a quality flag, the structure can be used to inform us about likelihood of the more dubious velocity assignments being the correct ones. For example, if a velocity assignment of 50.5 km$\,$s$^{-1}$ is given for a source with a poor quality flag, but its parent structure has a robust velocity assignment of 50.9\,km\,s$^{-1}$, then it is considered to be part of that same velocity group. If a poor velocity assignment is made to a source at a significantly different velocity to its neighbours within the hierarchy that have robust assignments, then we adopt the median velocity of those neighbours for that source. In this way, the velocity assignments for all sources (where possible) with a quality flag of either 1 or 2 were refined, and we adopt the quality flag of the most robust velocity within that group for all its constituent sources. We refer to the velocities refined in this way as $v_\mathrm{group}$. In Tab. \ref{tab:vflags} we present the distribution of quality flags assigned to the velocities determined and refined in this Section.

Finally, in order to ensure that velocity-coherent structures are retained, we perform a friends-of-friends grouping. Clusters of sources were identified by recursively linking groups that lie within a tolerance of 0\fdg06 in the $\ell$ and $b$ axes, and within 2.5\,\kms\ in line-of-sight velocities, $v_\mathrm{group}$, which approximates the characteristics of small (10\,pc-diameter) molecular clouds \citep[e.g.][]{Roman-Duval2010} at a distance of 5\,kpc -- the median near-kinematic distance for the revised source velocities along this sight-line. This allows, for example, the prominent large-scale filament that extends from $[\ell, b, v_\mathrm{group}] = [24\fdg16, 0\fdg40, 94\,$\kms$]$ to $[23\fdg57, 0\fdg59, 98\,$\kms$]$ to be identified as a single cluster. 2249 (92 per cent) of all GASTON sources are connected to one of the 211 clusters identified in this way, which contain up to 233 individual sources, with a median size of 25 sources. For each cluster, we also calculate the median group velocity, $v_\mathrm{cluster}$, and the associated standard deviation, $\sigma_\mathrm{cluster}$, using only the sources which have the most reliable velocity flags from that cluster.

The usage of 21 arcsec-resolution spectral line data with our 13 arcsec resolution element means that there is an element of beam-averaging that is likely to blend together velocity components that would be distinct at matching resolution. We have not made use of available survey data in $^{12}$CO (3--2) \citep{Dempsey2013} at 16.6 arcsec resolution,  due to optical depth considerations. While new facilities \citep[e.g.][]{Klaassen2019, Stanke2019} and instrumentation \citep[e.g.][]{Frayer2020} hold much promise, unbiased Galactic plane surveys in dense gas tracers such as N$_2$H$^+$, HCN, NH$_3$ or even C$^{18}$O at sufficiently high sensitivity and angular resolution are not currently available, and the FUGIN and RAMPS data present the best data for the moment. We note that CHIMPS2 \citep{Eden2020} aims to cover the GASTON GP field in $^{13}$CO (3--2) which, at an angular resolution of 15 arcsec and with a higher-critical density molecular gas tracer than that of the FUGIN, should complement GASTON upon its completion, though it is limited to $|b| < 0\fdg5$. We note that we do not use data from the Galactic Ring Survey \citep[GRS;][]{Jackson2006}, which have velocity resolution and sensitivity superior to the FUGIN data, because the higher angular resolution of the latter are more important in this case.

\subsection{Distance determination} \label{sec:distances}

Distances to sources lying within the Galactic plane are typically determined by measuring their radial velocity, and by assuming that the source is following a circular orbit described by a model of the Galactic rotation curve. For sources lying within the Solar Circle, each radial velocity corresponds to two kinematic distance solutions. In order to determine whether a source lies at the near- or far- kinematic distance, one either requires additional information, or some assumptions must be made. We determined kinematic distance solutions to each source using the Galactic rotation model of \citet{Reid2019} based on their $\ell, b, v_\mathrm{group}$ coordinates. As a first estimate, we use version 2.4.1 of the Bayesian distance calculator\footnote{\url{http://bessel.vlbi-astrometry.org/}} \citep{Reid2016}, adopting the recommended weightings for the priors based upon the spiral arm model, Galactic plane scale height, and proximity to sources with a measured parallax. This gives an independent distance estimate to each source (regardless of position within the hierarchy), but we note that this preferentially concentrates sources into the spiral arms. However, since the spiral structure of the Milky Way is very much an ongoing matter of research, we do not adopt these distance solutions directly. Rather, we use the distances calculated with these various priors to distinguish between the near and far kinematic distances. 

\begin{figure}
    \centering
    \includegraphics[width=\linewidth]{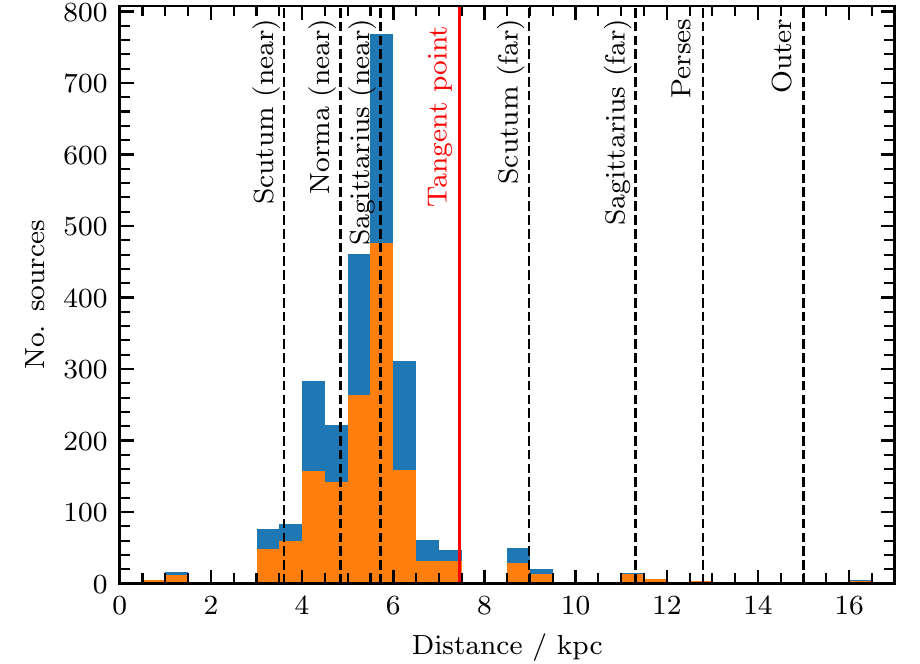}
    \caption{Distribution of kinematic distances in 0.5 kpc-wide bins, as determined for all sources (blue), and the subset of clumps (orange), in Sect. \ref{sec:distances}. The approximate locations of the intersection of the spiral arms with the line of sight, as in the \citet{Reid2019} model, are shown as vertical lines.}
    \label{fig:distances}
\end{figure}

2274 (93 per cent) of the 2446 sources are assigned to the near kinematic distance in this way, with only 120 (5 per cent) assigned to the far distance, and the remainder are located at the tangent point ($d \approx 7.5$ kpc, assuming the \citealt{Reid2019} distance to the Galactic centre of 8.15 kpc), where the near and far solutions are equal. By comparison, 77 per cent of the ATLASGAL clumps within the catalogue of \citet{Urquhart2018} have distances less than the tangent distance, after reducing the sample to those within the GASTON field, and adjusting for the difference in tangent distance associated with the adopted Galactic rotation model, and a similar figure of 79 per cent is found for 1.1 mm sources sources from the Bolocam Galactic Plane Survey \citep[BGPS;][]{Ellsworth-Bowers2015}. Only 35 per cent of clumps within the Hi-GAL catalogue of \citet{Elia2017} were assigned distances on the near-side of the tangent point. However, in the latter case, there is a bias towards far distances, as objects for which data (such as extinction data) that may allow a near determination to be made were unavailable are assumed to be located at the far distance. After removing clumps with this default assumption from the Hi-GAL catalogue, the number rises sharply to 89 per cent. We show the resulting distribution of distances in Fig. \ref{fig:distances}, and compare the values to their counterparts in other catalogues in Appendix \ref{app:distancecomparison}. We find that two-thirds of our distance estimates match with those of the other catalogues to within 1 kpc and can be considered to be in general agreement.

It is not surprising that a large fraction of sources are found to be located at the near distance as a result of the combination of Malmquist bias, and the fact that the $\ell=24\degr$ sight-line covers nearby sections of the Sagittarius, Scutum, and Norma spiral arms \citep{Reid2019}. However, we caution that the friends-of-friends analysis used here will inevitably lead to cases where sources at the far distance appear in clusters of sources for which our methodology favours a near distance.

In a manner similar to the determination of the cluster velocities, the median distance to each cluster is determined from the set of distances stemming from its constituent sources with the most robust velocity measurements, and then assigned to all sources within the cluster. Finally, we account for clusters which host maser sources with a known parallax distance. A total of 11 masers with parallax distances are located in the field, and after identifying the cluster associated with the $l, b, v$ position of each maser source, we assign the parallax distance to all sources within that cluster. There are a total of nine clusters that contain maser sources and are assigned distances in this way. Two of the eleven maser sources, G023.00-0.41 ($v = 79 \pm 5$\kms, $d = 4.88 \pm 0.36$~kpc) and G023.20-0.37 ($v = 82 \pm 10$\kms, $d=4.18 \pm 0.61$~kpc), are associated with objects within the same cluster, and since the associated distances are significantly different, a choice must be made. For this cluster, we adopt the parallax distance associated with G023.00-0.41, which is in stronger agreement with both the mean of the predetermined kinematic distances ($d = 5.20$~kpc) and systemic velocities ($77.5 \pm 0.2$\kms). A further two of the maser sources, G023.65-0.12 and G23.6, are located within the same cluster, but their distances and velocities are identical, and so present no conflict. The median discrepancy between the kinematic and maser parallax distances to these seven clusters is 0.41~kpc, though the largest difference is 1.73~kpc.


At this point, we have derived two distances for each object within the dendrogram, $d_\mathrm{group}$ and $d_\mathrm{cluster}$, both of which are derived principally from the sources' line-of-sight velocities, and the assumption that they simply follow circular motions about the Galactic centre. However, it is well known that line-of-sight streaming motions of up to $\sim 10$\kms\ can be present for molecular clouds, and peculiar motions within molecular cloud complexes further complicate this picture. To mitigate these effects in the subsequent analysis, the cluster distances were, therefore, used to determine any distance-dependent properties, and we note that the largest internal velocity dispersion of the clusters identified using the friends-of-friends grouping is 2.3\kms. The difference between the two sets of distances is minimal and, when considering the population of clumps, 98 per cent have a difference between $d_\mathrm{group}$ and $d_\mathrm{cluster}$ of less than 0.2 kpc.

\subsection{Physical properties} \label{sec:properties}

Estimating dust temperatures ($T_\mathrm{d})$ and column densities ($N_{\mathrm{H}_2}$) of molecular hydrogen from dust continuum imaging is typically done by fitting a modified blackbody to the spectral energy distribution (SED):

\begin{equation}
    I_\nu = \mu_{\mathrm{H}_2} m_\mathrm{H} N_{\mathrm{H}_2} \kappa_\nu B_\nu(T_\mathrm{d}),
\end{equation}

\noindent where $I_\nu$ is the specific intensity, $\mu_{\mathrm{H}_2}$ is the mean molecular weight per hydrogen molecular (with a value of 2.8), $m_\mathrm{H}$ is the mass of a hydrogen atom, $\kappa_\nu$ is the specific dust mass absorption coefficient, and $B_\nu$ is the Planck function evaluated at the frequency $\nu$ for the dust temperature. Three or more photometry points taken from imaging at submillimetre or millimetre wavelengths, convolved to the same resolution, are usually required (assuming both $\mu$ and $\kappa_\nu$ are fixed), providing a fit at the resolution of the longest wavelength being used -- typically 36 arcsec for \textit{Herschel} 500\,\micron\ data. However, the usage of dust temperatures derived at 36 arcsec-resolution for the 13 arcsec-resolution scales of the 1.15~mm data would mean that local minima and maxima would be beam-diluted. The different spatial frequencies probed by the ground- and space-based observatories also presents a further limitation.

To overcome these limitations, we adopt colour temperatures derived from the ratio of Hi-GAL 160 to 250\,\micron\ flux densities \citep[e.g.][]{Peretto2016, Rigby2018}, which have an effective resolution of 18 arcsec. The colour temperature is related to the ratio of flux densities in the following way:

\begin{equation} \label{eq:Tcol}
    \frac{S_{160}}{S_{260}} = \frac{B_{160}(T_\mathrm{col})}{B_{250}(T_\mathrm{col})} \left(\frac{250\,\upmu\mathrm{m}}{160\,\upmu\mathrm{m}}\right)^\beta.
\end{equation}

\noindent where $S_\nu$ is the source-integrated flux density $S_\nu = \int I_\nu d\Omega$. By adopting a fixed value of $\beta = 1.8$, and using 160 and 250~\micron\ photometry from Hi-GAL imaging, we sampled $T_\mathrm{col}$ from a grid of values ranging from 2.7 to 50 K, with 0.1~K intervals. Uncertainties on the \textit{Herschel} photometry result in uncertainties on $T_\mathrm{col}$ of $\sim$1.5~K. We first performed a convolution to effectively match the PSF of the 160~\micron\ image to that of the 250~\micron\ image following \citet{Aniano2011}, and both images were high-pass filtered to remove spatial frequencies larger than 6.5 arcmin using the {\sc nebuliser} application of the Cambridge Astronomy Survey Unit Tools software package\footnote{\url{http://casu.ast.cam.ac.uk/surveys-projects/software-release/background-filtering}} to approximate the spatial filtering present in the NIKA2 data. 

Many of the sources in the GP field consist of a mixture of compact and extended structure, and the flux densities of compact sources, therefore, usually contain contributions both of the compact object itself, and the extended emission structure on which they reside. We have therefore adopted the `clipping' scheme \citep{Rosolowsky2008} of flux estimation for $S_\nu$ (and thereby the mass) for each source, in which the local background level of flux is subtracted from the total, or put another way, only flux above the contour level defining the structure is counted in the integrated flux. Although \citet{Rosolowsky2008} prefer the `bijection' scheme, in which the entirety of the flux density integrated across the source's sky position is used for the mass calculation within molecular clouds, the clipping scheme is used here, since we are most interested in tracing the mass of the smallest structures within the hierarchy.

We determine the appropriate background level for each source by first performing a binary dilation on each of the source masks, which expands the perimeter by one pixel, and allows the encompassing perimeter pixels to be isolated (i.e. pixels just below the 1.15~mm contour level at which the source was identified). The background flux density at 1.15~mm is the contour level at which the source was separated from its parent structure, and was determined as the mean of the minimum pixel value within the source, and the maximum pixel value within the perimeter pixels. A slightly different approach must be taken to determine the background flux densities for the 160 and 250~\micron, because the morphology of emission at those wavelengths may differ from the morphology at 1.15~mm at which the source was defined. In contrast to the 1.15~mm case, some of the pixels within the boundary region may be brighter than some of those within the source area. At these wavelengths, the background value is therefore taken as half of the sum of the mean value of pixels within the perimeter, and the minimum pixel value within the source, thus approximating the 1.15~mm contour. 

The masses of the sources identified in the dendrogram are calculated from $S_{\nu}$, the monochromatic flux densities integrated over the solid angle subtended by the source at 1.15\,mm using

\begin{equation} \label{eq:mass}
    M = \frac{S_\nu ~ d^2}{\kappa_\nu B_\nu(T_\mathrm{col})},
\end{equation}

\noindent where the dust absorption coefficient is given by $\kappa_\nu = 0.1 (\nu / 1 \mathrm{THz})^\beta$ \citep{Beckwith1990}. We use a value evaluated at 1.15~mm (260 GHz) of $\kappa_{260} = 0.009$\,cm$^{2}$\,g$^{-1}$, assuming $\beta = 1.8$, and incorporating a gas-to-dust mass ratio of 100. This value is close to the value of 0.008 cm$^2$~g$^{-1}$ used in our previous NIKA study \citep{Rigby2018}, and that adopted by Hi-GAL in \citet{Elia2017}. We discuss the impact of different values of $\beta$ upon our results in Sect. \ref{sec:caveats}. For all sources, we use their cluster distance, $d_\mathrm{cluster}$, as determined in Sect. \ref{sec:distances}, and we decrease the integrated flux densities by 4 per cent to account for the median level of CO (2--1) contamination (determined in Appendix \ref{app:contamination}), adding a 4 per cent contribution to the corresponding uncertainties. Uncertainties are calculated using Monte Carlo methods for the primary quantities for each source (i.e., distance, absolute flux calibration uncertainty), and propagated to all derived quantities, such as mass and colour temperature. We note that the contribution of the factor of $\sim 2$ uncertainty on $\kappa_\nu$ is not included in our mass uncertainties since its effects are likely to be mostly systematic across the region, and will consequently not have an impact on the analyses presented here.

We note that, as a result of the 18 arcsec resolution of the colour temperature estimates, smaller-scale temperature variations will not be detected. This may result in the over-estimation of source masses where compact sources warmer than the beam-smeared average are present, and vise-versa for colder compact sources. At large radii, the temperature of protostellar sources is of the form $T(r) \propto r^{2/(4+\beta)}$, where $\beta$ is the dust emissivity spectral index \citep{Terebey1993} and, by adopting $\beta=1.8$, we can expect to underestimate dust temperatures of point sources at 18 arcsec by up to $\sim 11$ per cent compared to those at 13 arcsec. For resolved sources, this effect is unlikely to play a significant role, but we conservatively increase the uncertainties on $T_\mathrm{col}$ for all sources by 11 per cent to account for this effect.

We discuss the resulting distributions of mass and temperature in Sect. \ref{sec:newpopulation} and Sect. \ref{sec:evolution}, and present the overall distribution in Fig. \ref{fig:MTfull}.

\section{Results} \label{sec:results}

\begin{figure*}
    \centering
    \includegraphics[width=0.85\textwidth]{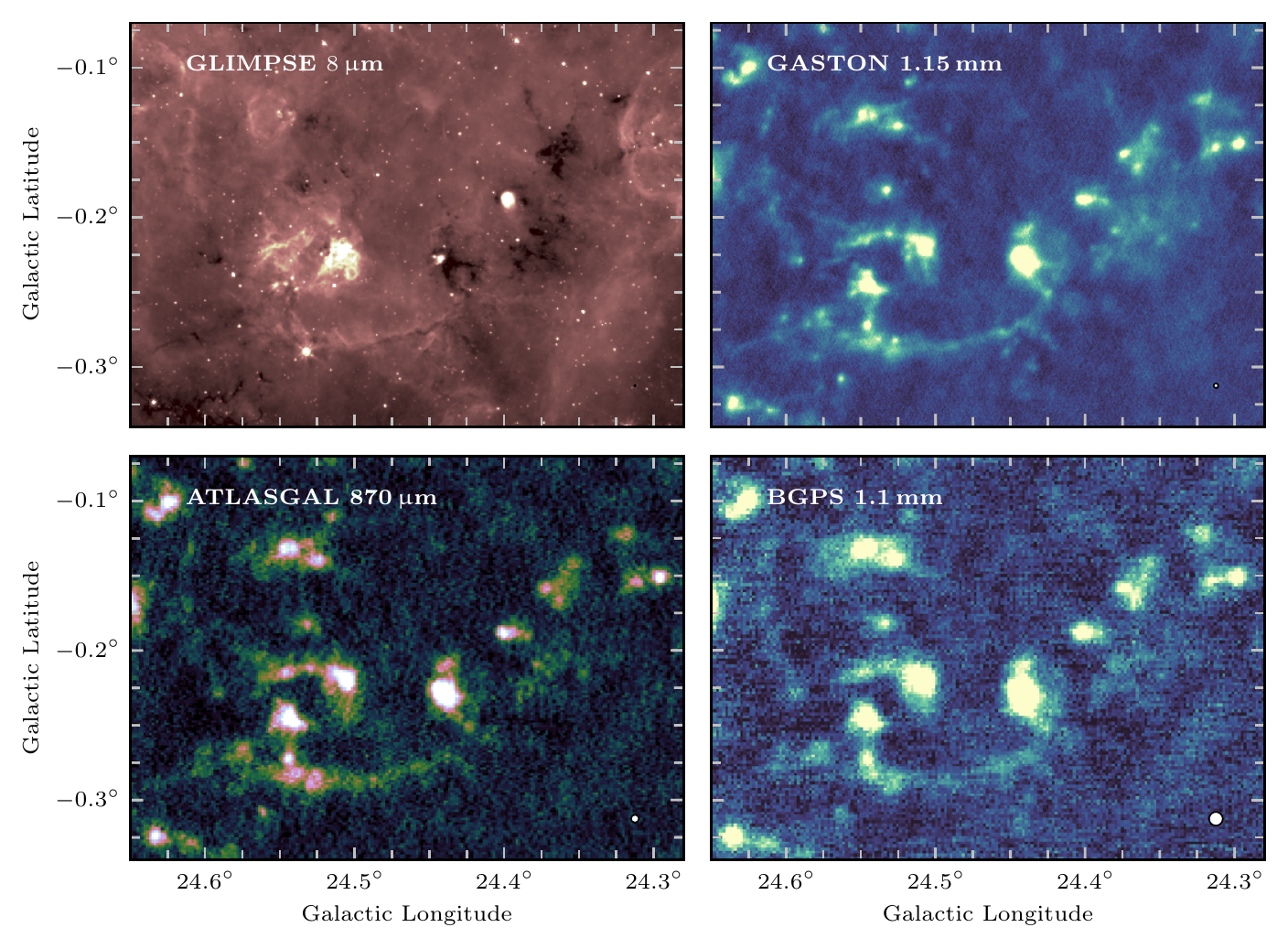}
    \caption{Comparison of the GASTON 1.15\,mm imaging (upper-right panel) to imaging from \textit{Spitzer}/GLIMPSE 8\,\micron\ \citep[upper-left;][]{Benjamin2003, Churchwell2009}, APEX/ATLASGAL 870\,\micron\ \citep[lower-left;][]{Schuller2009} and CSO/BGPS \citep[lower-right;][]{Ginsburg2013} imaging at 1.1\,mm. FWHM beam sizes are shown in the lower-right corners of each image. All images are shown at their native resolution and pixel sizes.}
    \label{fig:comparison}
\end{figure*}

\subsection{Comparison to other ground-based (sub-)millimetre data} \label{sec:comparison}

This particular region of the Galactic plane is well covered in many recent surveys of (sub-)millimetre continuum emission. The most similar ground-based surveys to our NIKA2 observations are the 1.1~mm BGPS at 33 arcsec resolution \citep{Aguirre2011, Ginsburg2013} and, at 870~\micron\ and with 19.2 arcsec resolution, ATLASGAL \citep{Schuller2009}. While the data from these surveys are very well matched to GASTON in terms of wavelength, they were designed to cover far wider regions of the GP (covering 192 and 360~deg$^2$, respectively), and so their sensitivities are lower than those of the data presented here.

We have identified a total of 1467 GASTON clumps (dendrogram leaves), compared to 346 ATLASGAL compact sources \citep{Urquhart2018} and 164 from the BGPS \citep{Ellsworth-Bowers2015} within the same sky area. Of the 164 sources from the BGPS, only 2 sources do not have centroid coordinates lying within three convolved beam radii\footnote{Our search parameters adopt convolved beam radii, that are the quadrature sum of the (smoothed) 13 arcsec GASTON beam and the beam size of the data set in question.} of a GASTON clump centroid. In both cases the BGPS source is located within a large-scale diffuse region in which the GASTON image has partially resolved and detected clumps around the periphery, and have centroid coordinates located far away from the particular region of emission extracted from the BGPS data. All of the ATLASGAL clumps lie within three convolved beam radii of a GASTON compact source, and they all fall within the boundaries of emission extracted by the dendrogram analysis in this work. The centroid coordinates of 1237 (84 per cent) and 1114 (75 per cent) out of the 1467 GASTON clumps are located further than three convolved beam radii of any BGPS or ATLASGAL source centroid, respectively. Although this positional matching based on centroid coordinates will be less effective for sources that are not centrally concentrated, $\sim$95 per cent of GASTON clumps have angular radii of less than three beam radii (i.e. 39 arcsec), and so it is clear that the majority of GASTON clumps are new identifications.

In Fig. \ref{fig:comparison}, we compare the GASTON 1.15~mm imaging in a region centred on $\ell = 24\fdg465, b=-0\fdg205$ to observations from ATLASGAL and the BGPS, and 8~\micron\ emission from the GLIMPSE. The greater angular resolution of the GASTON GP data compared to BGPS and ATLASGAL clearly allows new substructures to be resolved within previously-detected emission regions, in addition to the detection of new faint compact sources. By comparison to the \textit{Spitzer}/GLIMPSE 8~\micron\ image, it can also be seen that the 1.15\,mm continuum is effective at tracing structures seen in absorption as infrared-dark features as well as compact regions of bright emission at 8~\micron. Not all features from the 8~\micron\ image have discernible 1.15~mm counterparts, and although most of the faint and extended 8~\micron\ emission (presumably arising from hot dust grains or PAH emission at ionization fronts) is not seen in the GASTON image, there are visible counterparts to some of these features, such as the column extending from [$24\fdg62, -0\fdg10$] to [$24\fdg60, -0\fdg20$] and, surprisingly, the bow shock at [$24\fdg32, -0\fdg15$].

\begin{figure*}
    \centering
    \includegraphics[width=\textwidth]{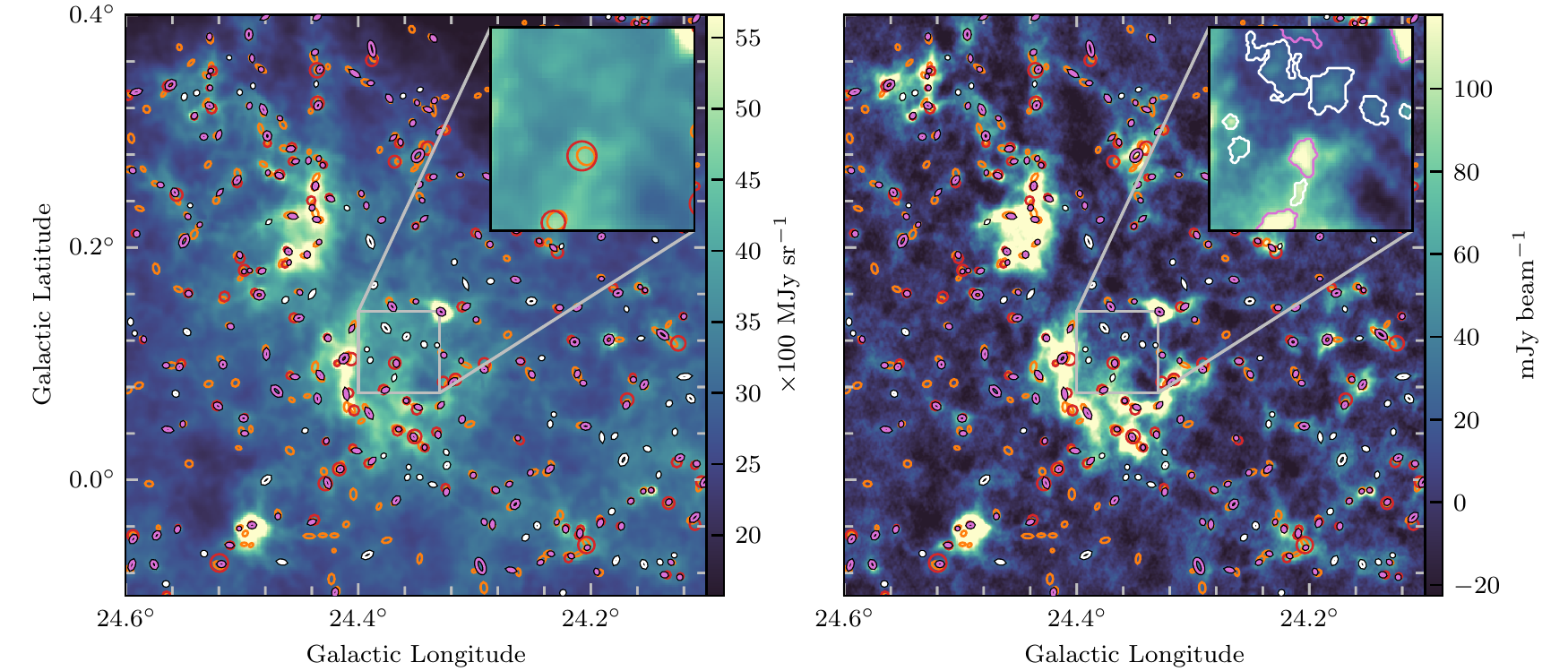}
    \caption{\textit{Herschel}/Hi-GAL \citep{Molinari2016} 250\,$\upmu$m (left panel) and NIKA2/GASTON 1.15\,mm (right panel) continuum emission maps for a $0.5\degr \times 0.5 \degr$ region centred on $\ell = 24.35\degr$, $b=0.15\degr$. The positions and sizes of the compact 250~\micron\ sources from the \citet{Molinari2016} Hi-GAL catalogue are shown as orange ellipses, and sources identified in the band-merged (BM) Hi-GAL catalogue of \citet{Elia2017} are shown as red circles. The positions of the 1.15\,mm GASTON clumps are shown by markers with black borders: those that have no cospatial 250\,$\upmu$m counterpart are given in white, while 1.15\,mm sources with counterparts are shown as pink ellipses. Both panels contain an inlaid zoomed region, with Herschel markers only shown on the left panel, while in the right pane1 -- for the sake of clarity -- we show the boundaries of the GASTON clumps, adopting the same colour scheme as for the ellipses in the main image.}
    \label{fig:cutout}
\end{figure*}

A more sensitive study of this region was recently carried out with the Large Millimetre Telescope at 1.1\,mm using AzTEC \citep{Heyer2018}. Those observations were centred on $\ell = 24\fdg5, b=0\fdg0$, with a radius of $0\fdg55$ -- partially overlapping with the GASTON GP field -- inside which they reached a median sensitivity of 9~mJy beam$^{-1}$ with a resolution of 8.5 arcsec. After rescaling to 1.15~mm, assuming a spectral energy distribution of the form $I_\nu \propto \nu^{3.8}$,  to match our NIKA2 observations, the point-source sensitivity would correspond to $\sim 7.6$~mJy beam$^{-1}$, making them around half as sensitive as those presented here. There are a total of 993 AzTEC compact sources within the region that overlaps with GASTON, after restricting the source catalogue to those considered by \citet{Heyer2018} to be robust, i.e. probability of a false-positive detection of $<2$ per cent, compared to 693 compact GASTON clumps, and 1175 GASTON sources of all kinds. 

698 (70 per cent) of the AzTEC source centroids fall upon a pixel associated with an extracted GASTON source, and the remaining 30 per cent are likely to be a result of: i) the greater resolution of the 32-m LMT Alfonso Serrano at a 1.1\,mm (8.5 arcsec) compared to the 30-m IRAM observations at 1.15\,mm (11.1 arcsec) and ii) the more pronounced spatial filtering in the AzTEC observations (removing spatial scales $> 50$ arcsec) compared to that of NIKA2 ($\sim 6\farcm5$); iii) the area of maximum sensitivity in the AzTEC map extends into regions of higher noise in the GASTON map, where our detection rate is lower. Furthermore, we have adopted dendrogram parameters in this study optimised to detect emission structures on all scales, and a treatment using harsher spatial filtering like that applied in the AzTEC data, or those of the ArT\'{e}MiS observations of \citet{Peretto2020} would be likely to increase the overall number of compact sources detected.

The statistics discussed in this Section are summarised in Tab. \ref{tab:matches}.

\begin{table*}
\centering
\caption{Summary of the results of catalogue cross-matching procedure used to compare the GASTON source statistics with other surveys in Sect. \ref{sec:comparison} and Sect. \ref{sec:newpopulation}. For each data set (i), we identify: (ii) the catalogue used; (iii) the wavelength; (iv) the angular resolution; (v) $N$, the total number of sources in the catalogue falling within the cropped GASTON GP field; (vi) $N_\mathrm{source}$, the number of source centroids that fall upon a pixel belonging to an identified GASTON source; (vii) $N_\mathrm{clump}$, the number of sources coincident$^1$ with a GASTON clump centroid; (viii) $N_\mathrm{source}$, the number of GASTON clump centroids coincident$^2$ with a catalogued source centroid; (ix) $N_\mathrm{unique}$, the number of GASTON clump centroids that are not coincident with a catalogued source centroid.}
\resizebox{\textwidth}{!}{%
\begingroup
\renewcommand{\arraystretch}{1.2} %
\begin{tabular}{lcccccccc}
\hline
Data & Reference & Wavelength & Resolution & $N$ & $N_\mathrm{source}$ & $N_\mathrm{clump}$ & $N_\mathrm{matched}$ & $N_\mathrm{unique}$ \\
(i) & (ii) & (iii) & (iv) & (v) & (vi) & (vii) & (viii) & (ix) \\
\hline
BGPS & \citet{Ellsworth-Bowers2015} & 1.1~mm & 33\arcsec & 164 & 162 & 162 & 230 & 1237 \\
ATLASGAL & \citet{Urquhart2018} & 870~\micron & 19.2\arcsec & 346 & 346 & 345 & 365 & 1102 \\
AzTEC & \citet{Heyer2018} & 1.1~mm & 8.5\arcsec & 993 & 698 & 420 & 336 & 357 \\
Hi-GAL & \citet{Elia2017} & BM$^3$ & 36\arcsec$^4$ & 857 & 731 & 734 & 702 & 765 \\
Hi-GAL & \citet{Molinari2016} & 70~\micron & 6\arcsec & 2807 & 1698 & 814 & 693 & 774 \\
Hi-GAL & \citet{Molinari2016} & 160~\micron & 12\arcsec & 4533 & 2334 & 1516 & 1145 & 322 \\
Hi-GAL & \citet{Molinari2016} & 250~\micron & 18\arcsec & 2591 & 1701 & 1466 & 1146 & 321 \\
Hi-GAL & \citet{Molinari2016} & 350~\micron & 24\arcsec & 1655 & 1247 & 1194 & 1075 & 392 \\
Hi-GAL & \citet{Molinari2016} & 500~\micron & 35\arcsec & 938 & 771 & 798 & 838 & 629 \\
\hline
\end{tabular}
 \endgroup}
    \begin{tablenotes}
      \small
      \item $^1$ Sources are regarded as coincident when the separation of their centroids is less than or equal to three convolved beam radii (Sect. \ref{sec:comparison}).
      \item $^2$ GASTON sources are regarded as coincident with a catalogued source as above with the exception of the Hi-GAL sources for which the circular or elliptical footprints are checked for overlapping GASTON sources (Sect. \ref{sec:newpopulation}).
      \item $^3$ Sources in this catalogue require at least three consecutive detections in the 160~\micron, 250~\micron, 350~\micron, and 500~\micron\ wavebands.
      \item $^4$ We adopt the most conservative beam size for the band-merged Hi-GAL catalogue.
    \end{tablenotes}
\label{tab:matches}
\end{table*}

\subsection{A new population of clumps} \label{sec:newpopulation}

NIKA2 is expected to be more sensitive to cold and compact dust sources within the Milky Way than \textit{Herschel}, owing to its greater angular resolution at longer wavelengths. To determine whether we have identified compact structures within the GASTON GP data that were not identified using the various extraction techniques applied to the \textit{Herschel} data, we compared our catalogue to two sets of Hi-GAL compact source catalogues. We performed a series of cross-matching exercises to identify any 1.15 mm GASTON clumps that have not been identified by either the five monochromatic source catalogues of \citet{Molinari2016}, or the `high-reliability' band-merged (BM) catalogue of \citet{Elia2017}. 

\begin{figure*}
    \centering
    \includegraphics[width=0.75\textwidth]{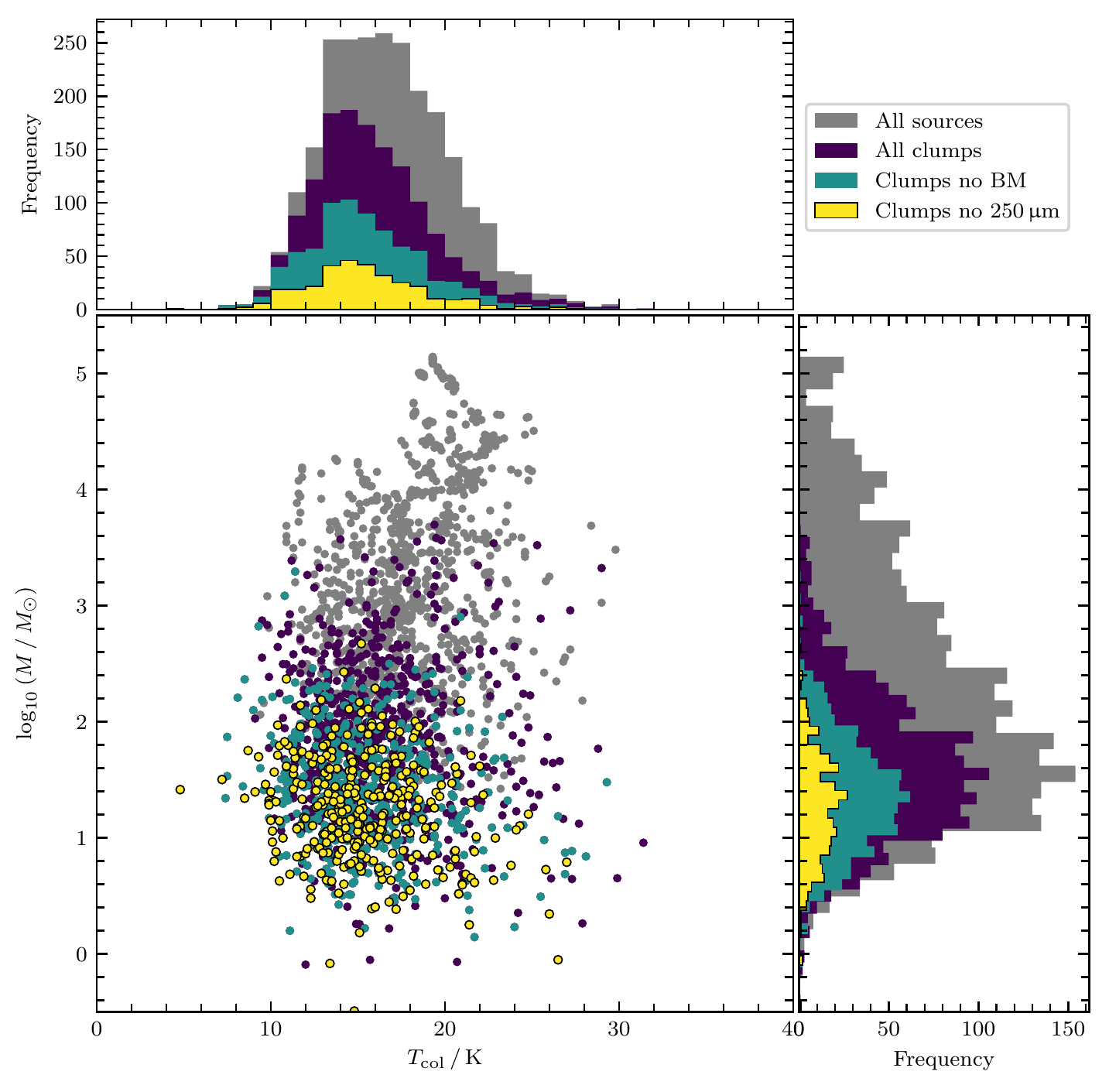}
    \caption{Mass and colour temperature distributions for all sources (grey), clumps (purple), clumps with no band-merged (BM) Hi-GAL counterpart (green), and clumps with no Hi-GAL 250\,\micron\ counterpart (yellow).}
    \label{fig:MTfull}
\end{figure*}

To carry out the source matching, we adopted the elliptical source sizes from the monochromatic catalogues, and the circular source sizes for the BM catalogue -- after convolving with a Gaussian kernel representative of the 1.15~mm NIKA2 beam -- to look for overlaps with the footprint of each dendrogram structure, and record the number of matches in each case (see Tab. \ref{tab:matches}). By comparison to the monochromatic catalogues, we found the lowest fraction of GASTON clumps that do not match to any Hi-GAL compact source in the 250\,\micron\ band, with 321 GASTON clumps (22 per cent) without a 250\,\micron\ counterpart, and 765 (40 per cent) with no band-merged counterpart. We compare the \textit{Herschel} 250~\micron\ data and the GASTON GP 1.15~mm data for a $0\fdg5 \times 0\fdg5$ region centred on $\ell = 24\fdg35, b=0\fdg15$ in Fig. \ref{fig:cutout}, overlaying markers representing catalogued sources from the 1.15~mm dendrogram extraction and the two Hi-GAL catalogues. The highest match fraction with sources in the 250\,\micron\ was expected, as these particular \textit{Herschel} observations represent a compromise between the greater angular resolution available at the shorter wavelengths, and the greater sensitivity to the coldest structures at the longer wavelengths. Although the 500\,\micron\ catalogue represents the closest match in wavelength, the large difference in angular resolution results in a substantially different segmentation of the emission under the different source extraction methods.

In Fig. \ref{fig:MTfull}, we compare the distributions of mass and colour temperature, $T_\mathrm{col}$, for the full sample of sources. Their statistical properties are summarised in Tab. \ref{tab:MT}. The new clumps identified by our NIKA2 observations with no counterpart in the Hi-GAL catalogues fall, on average, towards lower mass and temperature than those associated with \textit{Herschel}-catalogued sources. We compare the $\log_{10}(M/M_\odot)$ and $T_\mathrm{col}$ distributions for the clumps that do have 250~\micron\ Hi-GAL matches and those that do not, using two-sample Anderson-Darling. The results show that the distributions of both properties in the two populations are inconsistent at a high confidence level, with $p$-values of $< 0.02$ in both cases.

There are 14 clumps with masses in excess of $100$\,M$_\odot$ that are new detections. We note that to compare these values to other survey data, such as ATLASGAL, one must consider the masses measured in the same way, and the ATLASGAL clump masses of \citet{Urquhart2018} adopt a method more similar to the `bijection' scheme (as discussed in Sect. \ref{sec:properties}). The approximate conversion between the `clipped' masses reported in this study, and the equivalent bijected masses, calculated from combined \textit{Herschel} and NIKA2 column density maps \citep[following the method of][]{Rigby2018} is $M_\mathrm{bijection} \approx 41 \times M^{0.67}$, with a simple linear least-squares fit in log-space. In this metric, 33 of the 321 new clumps have $M_\mathrm{bijection} > 1000~M_\odot$, and may be considered as high-mass. 256 of the 321 new clumps (80 per cent) have no compact 70~\micron\ counterpart found by matching to the \citet{Molinari2016} catalogue, and so may be considered as candidates for starless sources, though we caution that this is an upper limit, and a lack of a 70~\micron\ match does not guarantee a lack of embedded star formation \citep[e.g.][]{Traficante2017}.

\begin{table}
    \centering
    \begin{threeparttable}
    \caption{Statistical results -- mean, standard deviation, sample size, standard error on the mean -- of the masses and colour temperatures for the various samples presented in Fig. \ref{fig:MTfull}.}
    \begin{tabular}{l l l r l}
    \hline
    Sample & Mean & $\sigma$ & $N^1$ & $\sigma/\sqrt{N}$ \\
    \hline
    $\log_{10}(M/M_\odot)$\\
    All sources           & 2.24  & 1.08  & 2443  & 0.02 \\
    Clumps                & 1.60  & 0.63  & 1464  & 0.02 \\
    Clumps no BM          & 1.35  & 0.49  & 763  & 0.02 \\
    Clumps no 250~\micron & 1.26  & 0.46  & 319  & 0.03 \\
    \hline
    $T_\mathrm{col} / \mathrm{K}$\\
    All sources           & 16.66 & 3.59 & 2443 & 0.07 \\
    Clumps                & 15.84 & 3.56 & 1464 & 0.09 \\
    Clumps no BM          & 15.45 & 3.53 & 763 & 0.13 \\
    Clumps no 250~\micron & 15.35 & 3.42 & 319 & 0.19\\
    \hline
    \end{tabular}
    \begin{tablenotes}
    \label{tab:MT}
      \small
      \item $^1$ Note that 3 clumps have 160/250~\micron\ flux ratios that correspond to colour temperatures that outside of the range $2.7 \ge T_\mathrm{col} \le 50$~K, and so have no colour temperatures or masses derived here.
    \end{tablenotes}
  \end{threeparttable}
\end{table}

We point out here that the various catalogues used in the cross-matching exercises in this Section and in the previous one, were generated using different source extraction techniques. We caution that we expect the total number of sources in the different data sets to vary with the methodology, but it is clear that we detect a substantial number of previously undetected sources, with the 321 sources that do not coincide with 250~\micron\ compact sources representing the most conservative estimate.

\subsection{The evolution of star-forming clumps} \label{sec:evolution}

To characterise our sample, we have computed the infrared-bright fraction for each source, $\firb$, based on GLIMPSE 8~\micron\ imaging \citep{Benjamin2003, Churchwell2009}. The infrared-bright fraction is defined as the fraction of pixels within the source boundary that are brighter than the median value of all pixels within a 4.8-arcmin box centred on each pixel (see Watkins et al. \textit{in prep} for a full description). The filter size was selected so that the most extended sources within the 8~\micron\ image -- typically bubbles associated with \ion{H}{ii} regions -- could be compared to their local background. Although this filter scale represents a different spatial scale for the nearest and farthest sources within the region, it offers robust measurements for sources with angular sizes less than half of the filter scale, which is true for all GASTON clumps. We note that a very similar the classification scheme was used by \citet{Battersby2011}, who used GLIMPSE 8~\micron\ images to classify a sample of Hi-GAL sources.

In Sect. \ref{sec:newpopulation}, each clump was also matched to the 70\,\micron\ catalogue of \citet{Molinari2016} and we record the sum of the positionally-matched 70\,\micron\ source integrated flux densities. The 70\,\micron\ luminosity is known to be a good tracer of the bolometric luminosity of any embedded sources \citep[e.g.][]{Dunham2008, Ragan2012}. We calculate the bolometric luminosity by using the \citet{Elia2017} relationships:
\begin{equation}
    \frac{L_\mathrm{bol}}{L_\odot} = \begin{cases} 2.56 \left(\frac{S_{70}}{\mathrm{Jy}}\right)^{1.00}  \left(\frac{d}{\mathrm{kpc}}\right)^2 & \mbox{if } S_{70} \ge 50 \, \mathrm{Jy}, \\  3.29 \left(\frac{S_{70}}{\mathrm{Jy}}\right)^{0.79}  \left(\frac{d}{\mathrm{kpc}}\right)^2 & \mbox{if } S_{70} < 50 \, \mathrm{Jy}, \end{cases}
\end{equation}

\noindent where $S_{70}$ is the sum of the integrated flux densities of any compact sources from the catalogue of \citet{Molinari2016} that lie within one beam-radius of a GASTON clump boundary.

\begin{figure}
    \centering
    \includegraphics[width=\linewidth]{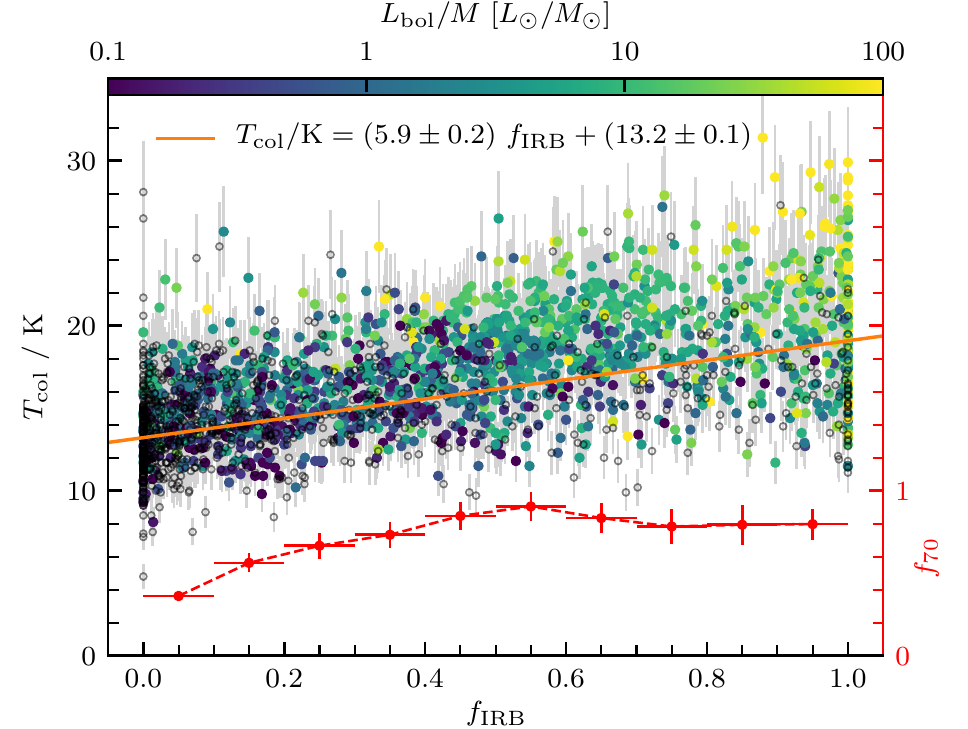}
    \caption{Colour temperature as a function of the infrared-bright fraction, $\firb$. The coloured points show the relation for all GASTON sources which have at least one matched Hi-GAL 70~\micron\ compact source from \citet{Molinari2016}, and are colour-coded by the ratio of their bolometric luminosity to 1.15~mm mass. The empty circles show the relationship for unmatched GASTON sources. The trend determined by a linear least-squares fit is shown as the solid orange line. The red points show, on the secondary $y$-axis, the fraction of GASTON sources that are associated with compact 70~\micron\ sources as a function of $\firb$ in 0.1-wide bins.}
    \label{fig:fIRB}
\end{figure}

The infrared-bright fraction, $\firb$, may serve as an indicator for the evolutionary stage of a source: sources that are completely infrared-dark ($\firb = 0$) are likely to be at the earliest stages of evolution, containing no observable indications of star-formation in terms of their 8\,\micron\ flux and, conversely, sources that are completely infrared-bright ($\firb = 1$) are likely to be sources exhibiting signs of advanced star-formation, most commonly in the form of emission from young stellar objects, \ion{H}{ii} regions, or heated polycylcic aromatic hydrocarbon (PAH) molecules. Although a more detailed analysis will be presented in Watkins et al. (\textit{in prep.}), we demonstrate the basic utility of $\firb$ as an evolutionary indicator in several ways in Fig. \ref{fig:fIRB}. Firstly, there is a positive correlation (Pearson correlation coefficient $\rho = 0.57$, $p\mathrm{-value} \ll 0.1$) between $\firb$ and the source temperature as traced by $T_\mathrm{col}$. Secondly, there is also a positive correlation ($\rho = 0.57$, $p\mathrm{-value} \ll 0.1$) between $\firb$ and $\log_{10}(L_\mathrm{bol}/M)$. Temperature and $L_\mathrm{bol}/M$  are both generally regarded as robust tracers of the time-evolution of clumps \citep[e.g.][]{Urquhart2014, Urquhart2018, Molinari2016a, Svoboda2016, Elia2017}. Finally, we show that the fraction of GASTON sources that are associated with compact 70~\micron\ sources from Hi-GAL \citep{Molinari2016} also increases with $\firb$, and reaches a plateau of $\sim85$ per cent for sources with $\firb \gtrsim 0.5$.

It is important to be aware of the potential for distance biases using the $\firb$ parameter, since darkness at 8~\micron\ usually requires a bright background against which the radiation can be absorbed. We therefore reduce our sample in the following analysis to include only those clumps that lie at distances between 3.5 and 7.0~kpc. This criterion serves a dual function; firstly, it limits the effect of the distance bias in $\firb$ since infrared-dark clouds are generally located at the near kinematic distance \citep[e.g.][]{Ellsworth-Bowers2013}, and the tangent for kinematic distances at this longitude range is $\sim 7.5$~kpc. Secondly, this limits the role of a varying spatial resolution element in the source extraction process, and these particular limits mean that the most distant sources have only a factor of 2 lower spatial resolution than the nearest ones. Conveniently, these limits remove only $\sim$10 per cent of the sample.

We define four categories through which we divide the sample of clumps based upon their infrared-bright fractions that can be used to broadly trace evolutionary phase: i) $\firb < 0.040$ for predominantly infrared-dark clumps; ii) $0.040 \ge \firb < 0.229$ for clumps that are mostly infrared-dark; iii) $0.229 \ge \firb < 0.663$ for clumps that are partially infrared-bright; iv) $\firb \ge 0.663$ for clumps that are predominantly infrared-bright. The specific values of $\firb$ were chosen such that the four sub-samples are of equal size, each containing a total of 321 clumps. The fraction of clumps in each category that are associated with compact 70~\micron\ counterparts, $f_{70}$ also increases monotonically with the proposed evolutionary sequence, with 21, 33, 59 and 69 per cent of the clumps having 70~\micron\ counterparts in samples i)--iv), respectively.
 
Tracing the precise time-evolution of star-forming regions is not a trivial matter, but although we caution that $\firb$ does not give an \textit{absolute} value of `evolutionary stage' or age for individual sources, it may be regarded as \textit{relative} age indicator and, therefore, a suitable measurement when dealing with large samples of sources. Using a similar methodology to that used in this paper, \citet{Battersby2017} estimated the fraction of a sample of dense molecular regions above a column density threshold associated with high-mass star formation that were starless or star-forming. By comparing the relative fractions of these categories to those associated with methanol masers and ultra-compact \ion{H}{ii} regions -- which have established lifetimes -- the authors estimated the lifetime of the starless and star-forming phases as 0.2--1.7 and 0.1--0.7~Myr, respectively. These two timescales may broadly map onto stages i)--ii) (predominantly starless) and iii)--iv) (predominantly star-forming), respectively, indicating a total duration of $\lesssim 2.5$~Myr for the evolution from $\firb = 0$ to $\firb = 1$.

\begin{figure*}
    \centering
    \includegraphics[width=\textwidth]{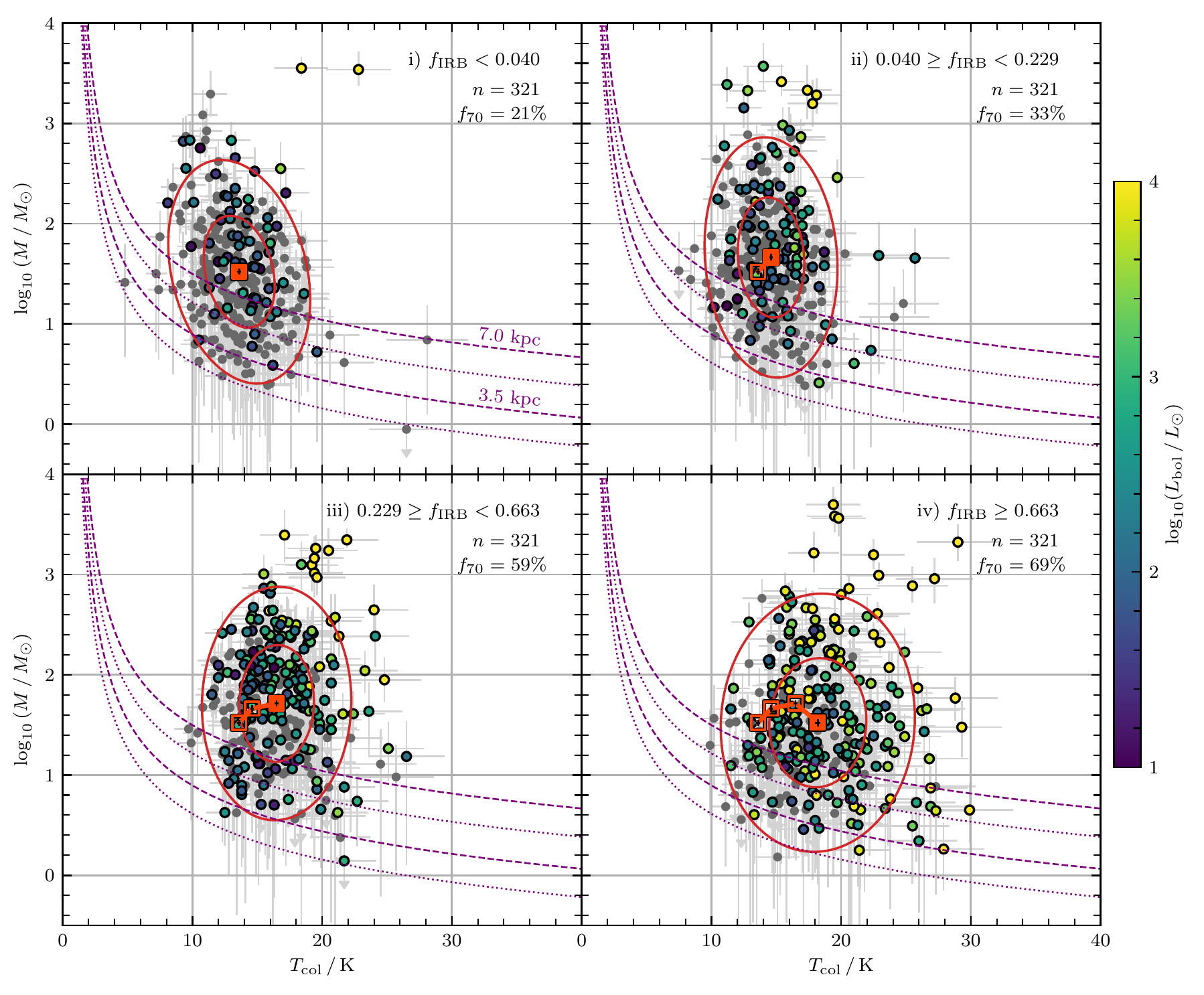}
    \caption{Mass--colour temperature relationships for the four subsamples of GASTON clumps, classified by their infrared-bright fraction, $\firb$. Masses are calculated using the `clipping' method. Filled orange squares show the mean value of the quantities plotted in each axis, while the open squares show the mean values for the previous stages in the proposed evolutionary sequence. In each case, the standard error on the mean in both axes is shown in black within the marker shapes. The colour of the points shows the total bolometric luminosity from matched 70~\micron\ compact sources, while the grey points have no compact 70~\micron\ counterparts from \citet{Molinari2016}. Red ellipses show the approximate 1-$\sigma$ and 2-$\sigma$ confidence intervals for the distribution, containing $\sim$68 and 95 per cent of the data points, respectively. The dashed and dotted purple curves correspond to the 1.15~mm point-like and extended source sensitivities for the combination of the data and source extraction parameters at the near- and far-distance limits of the sample.}
    \label{fig:evolution}
\end{figure*}

In Fig. \ref{fig:evolution}, we display these four samples in (logarithmic) mass and colour temperature space. By plotting the mean position of the four sub-samples in the two axes, a progressive movement of the distributions is seen as the clumps get brighter at 8~\micron. These mean values suggest that the clumps, on average, gain a moderate amount of mass -- increasing by $\sim 0.2$ dex between stages i) and iii), reflecting a 60 per cent increase -- before losing the same amount of mass, all the while increasing in temperature. The errors on the mean values indicate that although these changes in mass are subtle, they are significant at the 3.1$\sigma$ level between stages i) and ii), at the 4.3$\sigma$ level between stages i) and iii), and at the 4.0$\sigma$ level of significance between stages iii) and iv). The increase in source counts as the GASTON GP project progresses will allow these values to be revisited with greater statistical power. The mean values and their associated standard errors are given in Tab. \ref{tab:MTvalues}. A further striking feature of Fig. \ref{fig:evolution} is the increasing spread in temperatures along the evolutionary sequence. This behaviour is expected as a consequence of the more rapid evolution of the highest-mass sources, in addition to the mass function. The more numerous low-mass sources are expected to evolve relatively slowly, only increasing in temperature, and losing mass at later stages as the clump is dispersed. The rarer high-mass clumps are expected to rapidly gain in mass and temperature at early sages, before losing mass at later stages.

\begin{table}
\centering
\caption{Mean values and associated standard errors of the distributions of the four evolutionary stages presented in Fig. \ref{fig:evolution}.}
\begin{tabular}{@{\extracolsep{6pt}}ccccc@{}}
\hline \hline
\noalign{\smallskip}
Stage & \multicolumn{2}{c}{$\log_{10}(M/M_\odot)$} & \multicolumn{2}{c}{$T_\mathrm{col}$ / K}  \\ 
\noalign{\smallskip}
\cline{2-3} \cline{4-5}
\noalign{\smallskip}
& Mean & Error & Mean & Error \\
\noalign{\smallskip}
\hline
\noalign{\smallskip}
i) & 1.52 & 0.03 & 13.61 & 0.15 \\
ii) & 1.66 & 0.03 & 14.60 & 0.14 \\
iii) & 1.71 & 0.03 & 16.49 & 0.16 \\
iv) & 1.52 & 0.04 & 18.20 & 0.21 \\
\noalign{\smallskip}
\hline
\end{tabular}
\label{tab:MTvalues}
\end{table}
 
 This reflects the behaviour that might be expected from a clump-fed model of star formation, whereby the mass gain of the clump is the result of accretion from its direct environment, and the clump is subsequently dispersed by feedback as star formation progresses. By contrast, in a core-fed scenario, such clumps do not exhibit an initial increase in mass, since all of the mass is in situ before star-formation begins, which seems incompatible with Fig. \ref{fig:evolution}. Furthermore, the movement of the mean points in Fig. \ref{fig:evolution} bears a striking resemblance to the mass-temperature tracks described by the clump-fed models of \citet[][see their Fig. 7]{Peretto2020}, supporting the coeval mass growth of clumps and embedded protostars within. However, it is important to note that the sizes of the GASTON sources presented here are, on average, larger than those presented in \citet{Peretto2020}. This might suggest that the same accretion process from large to small scales occur over a relatively wide range of scales. A qualitatively similar behaviour as that seen in Fig.~ \ref{fig:evolution} is observed  by \citet[][see their Fig. 22]{Elia2017}. In that study, clumps were categorised as starless or star-forming based upon the presence of 70~\micron\ emission, with the star-forming sample further divided according to the detection or non-detection of 21--24~\micron\ counterparts, and coincidence with \ion{H}{ii} regions. 
 
 We highlight that each of the four subsamples is likely to contain a mix of both core-fed and clump-fed sources, and so the signal present here in the progressive movement of the mean position will be moderated by the core-fed sources toward the low-mass end of the mass distribution. If the clump-fed scenario is more likely to be important for the highest-mass clumps, then this behaviour could be reflected in the upper envelope of the distributions. The 95th percentile in mass changes by $+0.08$, $-0.03$, and $+0.06$ dex, while the 95th percentile in temperature increases by 0.7~K, 2.7~K and 4.4~K between stages i)--ii), ii)--iii), and iii)--iv), respectively. The lack of a reduction in the 95th percentile in logarithmic mass in the final stages is more difficult to interpret, as it suggests that the clump dispersal phase of the most massive clumps is either not being sampled well here, or lasts a relatively long time. The former explanation could easily be true, as the number of sources in the 95th percentile is only 16 in each stage, and thus the measurement suffers from small numbers. \citet{Urquhart2014} argued that, although the accretion phase for high-mass protostars is probably very rapid, they may then take a relatively long time to begin dispersing the enveloping clump, resulting in the clustering of sources at the apex of their mass-luminosity evolution. Qualitatively, this would result in the same behaviour seen here with these highest-mass clumps spending much of their time in the high-mass, high-temperature part of this diagram. This is also supported by the clump-fed models of \citet{Peretto2020}, whose evolutionary tracks for the highest-mass stars shows an accretion phase that is relatively short-lived compared to the lifetime approaching their maximum mass.
 
 It is unlikely that this sequence arises as a result of observational effects such as sensitivity or completeness. The point-source sensitivity limits are displayed in each panel of Fig. \ref{fig:evolution}, and do not display any signs of correlation with the movement of the mean position of the distributions in stages i)--iii); lower-mass clumps are more likely to be observable at later stages when temperatures are higher, and so this effect may in fact be suppressing mean value of the mass in the middle stages, and the average mass gain in high-mass sources is therefore likely to be greater than the value of 60 per cent deduced from this study.
 
 The source extraction, which was carried out on the 1.15~mm map, is completely independent of the $\firb$ parameter, and so we do not expect any systematic bias in terms of completeness as a function of evolutionary stage. The effects of mass completeness manifest themselves in this parameter space by the position of the peak of the distribution. The real distribution of clump masses is expected to consist of a power law $N \propto M^\alpha$, where $\alpha < 0$, and so the turnover close to the mean value reflects the completeness that is a function of the sensitivity and source extraction methodology. Since both of these factors are consistent across all samples, it is difficult to conceive a scenario in which differences in completeness could be dominating the position of the mean. 
 
\subsection{Caveats on the proposed evolutionary scenario} \label{sec:caveats}

In the previous Section we have provided evidence, acquired from the catalogue of clumps identified at 1.15\,mm, that the overall population is most massive at intermediate evolutionary stages, as traced by the $\firb$ proxy for relative evolution. In this Section, we explore various caveats to this analysis, and test the robustness of the observed trend.

\begin{figure}
    \centering
    \includegraphics[width=\linewidth]{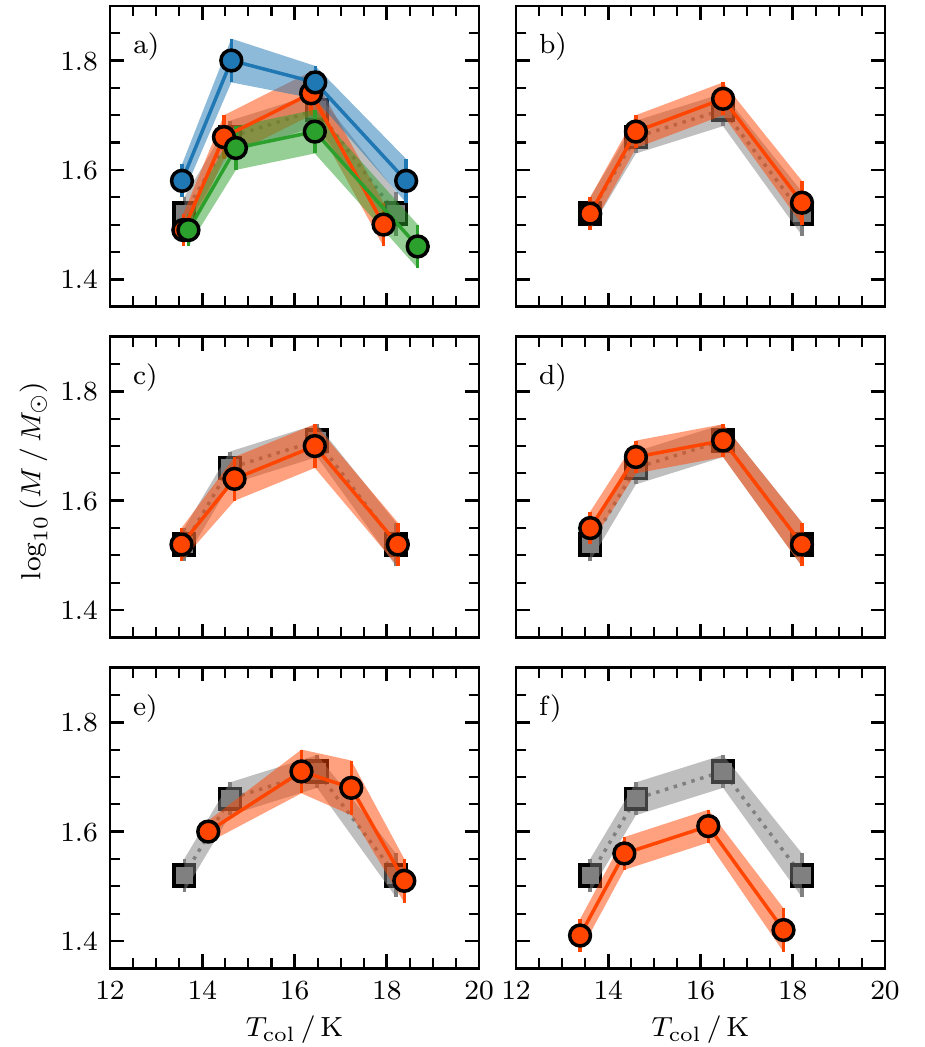}
    \caption{Mean evolutionary trend after altering the samples to: a) displace a randomly-selected third of clumps to the far kinematic distance (using three sets); b) adopt $d_\mathrm{group}$ as the distance estimate; c) adjust the distance limits to $4 < d_\mathrm{cluster} < 6$ kpc; d) normalise the clump masses to the mean distance of 5.35 kpc; e) adopt equally-spaced $\firb$ limits for the subsamples; f) use alternative values of $\beta$. In each case, the original trend is shown as grey squares, with mean positions of the altered samples overlaid in coloured circles. The shaded regions show the $\pm 1\sigma$ confidence limits.}
    \label{fig:MTcomparison}
\end{figure}

Firstly, our methods for acquiring clump distances in Sect. \ref{sec:distances} have resulted in a large fraction -- 93 per cent -- of sources that are located at the near kinematic distance. We expect a high fraction along this line of sight, and it is, perhaps, not surprising that a higher fraction is recovered than for clumps in ATLASGAL (77 per cent), BGPS (79 per cent), or Hi-GAL (89 per cent of the `good-quality' assignments), due simply to the greater mass sensitivity of GASTON, and the greater prevalence of low-mass compared to high-mass clumps. However, we have tested the robustness of the proposed evolutionary trend by repeating the analysis three times after first randomly assigning 1/3 of the population of clumps to their far kinematic distance solutions. The resulting trends are illustrated in Fig. \ref{fig:MTcomparison}, and the overall effect is, in two of the three cases, to slightly reduce the significance of the differences between stages i)--ii), i)--iii), and iii)-iv) compared to the original distance assignments, while the significance is increased in the third case. This tendency to slightly reduce the significance can be mainly ascribed to the resulting smaller sample sizes, as the clumps now assigned to the far-kinematic distance fall outside the distance limits of the sample, consequently increasing the error on the mean.

Another choice affecting the results was the decision to adopt the cluster-averaged distances, $d_\mathrm{cluster}$, as opposed to the dendrogram group-averaged distances, $d_\mathrm{group}$. If we adopt the latter in place of the former, and repeat the analysis, we see almost no change in the results, since the average difference in the two distances in on the order of one hundred parsecs. In fact, as can be seen in panel b) of Fig. \ref{fig:MTcomparison}, the evolutionary trend is slightly strengthened in this scenario with the largest change coming between stages i) and iii), now at the 4.5-$\sigma$ level of significance, compared to 4.3-$\sigma$ in the original analysis.

The choice of distance limits used could also have an effect upon the recovered trend, and we repeated the experiment by adopting more conservative distance limits, such that a maximum factor of 1.5 difference in the spatial resolution at the near and far cut-off was used, i.e. $4 < d_\mathrm{cluster} < 6$ kpc. The effect here is, again, to reduce the sample size, now with 259 or 260 clumps in each stage compared to the original 321, and thereby reduce the significance of the trend. In this case, the largest change in $\log_{10}(M/M_\odot)$ between stages i) and iii) is still significant at the 3.7-$\sigma$ level (panel c) of Fig. \ref{fig:MTcomparison}). 

\begin{figure}
    \centering
    \includegraphics[width=\linewidth]{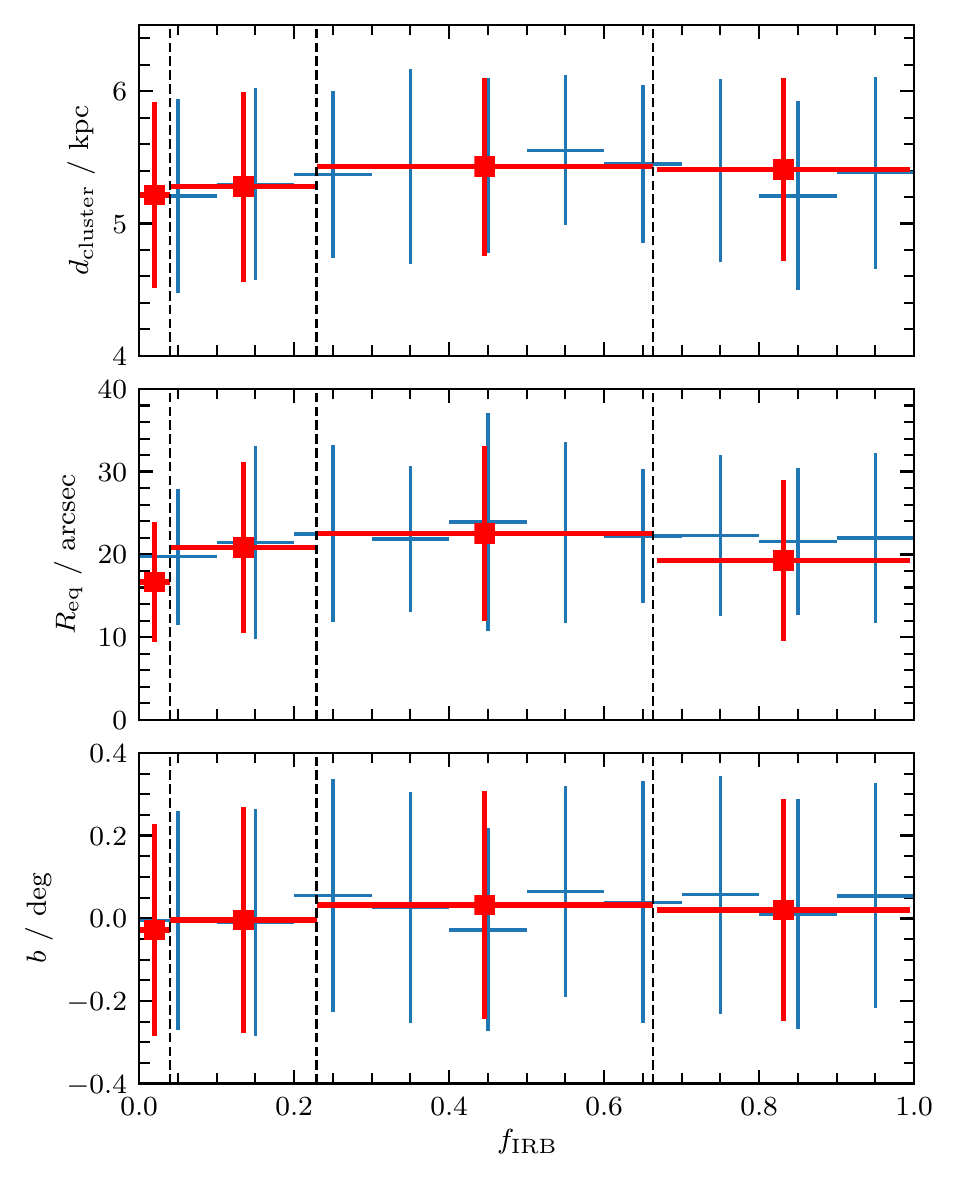}
    \caption{Average values of distance (top), angular size (middle) and centroid latitude (bottom) of GASTON clumps as a function of the infrared-bright fraction. The red data points show the mean values for samples i)--iv), and the blue data points show the mean values for clumps sampled in 0.1-wide bins of $\firb$. The vertical error bars indicate the standard deviation of the populations, and the horizontal error bars indicate the $\firb$ range of the sample. Dashed vertical lines indicate the boundaries used for the samples i)--iv).}
    \label{fig:biases}
\end{figure}

We also note the presence of potential biases in both distance and angular size as a function of $\firb$. Relative to one another, the average values of both quantities shows the same behaviour as the average mass throughout the four proposed evolutionary stages. A bias towards nearer distances for the more infrared-dark sources can be understood by their requirement for a bright infrared background against which they can be seen, though it is not clear why there should be a bias for the brightest sources to also be located closer to the observer. The bias towards larger sources at intermediate values of $\firb$ can be understood by the nature of the median filter, by which sources should tend towards $\firb = 0.5$ as their size approaches the filter size, by construction. We illustrate these biases in Fig. \ref{fig:biases}, indicating the mean values and standard deviations. One way to explore the effect of the distance bias is to normalise all of the clump masses to the same distance, and we do this by rescaling the masses as if they were located at the mean distance within the sample, multiplying them by a factor of $(5.35\,\mathrm{kpc} / d_\mathrm{cluster})^2$, and we include the modified evolutionary trend in panel d) of Fig. \ref{fig:MTcomparison}. The bias in distance is extremely mild, and has little effect upon our conclusions. However, the effect of the bias in angular size is more difficult to test, and so we caution that there may be an effect at play here though, again, the effect is probably very mild, considering the overlapping distributions. In Fig. \ref{fig:biases}, we also show the lack of any significant bias in $\firb$ as result of the latitude of the clumps, indicating that this particular line of sight has a sufficiently bright infrared-background so as to make any latitude-dependent bias negligible.

In Sect. \ref{sec:evolution}, the limits to $\firb$ were selected to equalise the size of the four samples, and thereby eliminate the statistical effects of varying sample sizes. Since $\firb$ is expected to trace the relative evolution of each clump (however non-linearly), a secondary effect of this choice of samples is that a similar fraction of the clumps' total lifetimes may be covered in each stage. We can, however, investigate whether these particular boundaries of $\firb$ that define the evolutionary stages, are responsible for the apparent sequence. Therefore, we conducted a further test, by defining the subsamples with equally-spaced $\firb$ limits at $\firb = 0.25, 0.50, 0.75$. The four samples i)--iv) now contain 656, 202, 148, and 248 clumps, and, again, the same general trend is present (panel e) of Fig. \ref{fig:MTcomparison}), though at a reduced significance, with 2.35-, 1.57- and 2.73-$\sigma$ changes in the value of $\log_{10}(M)$ between stages i)--ii), i)--iii), and iii)--iv), respectively. that equal time is spent -- on average -- in each stage. It is conceivable that other $\firb$ limits could be adopted to either amplify or diminish the significance of the proposed evolutionary tend, but with a lack of physical or statistical justification, we explore these no further here.

Our calculation of temperatures and masses in Sect. \ref{sec:properties}, could also be having an effect upon the evolutionary trend. We adopted a value of $\beta=1.8$, based upon the Galactic plane average value, as measured by \textit{Planck} \citep{PlanckCollaboration2011}, and thereby calculated a value of $\kappa_{260}$. However, the value of $\beta$ has been shown to vary as a function of frequency \citep[e.g.][]{PlanckCollaboration:2014}, and so we test the impact of a two-valued $\beta$ upon the reported evolutionary trend. Following \citet{PlanckCollaboration:2014}, we adopt a far-infrared value of $\beta_\mathrm{FIR} = 1.88 \pm 0.08$ for our calculation of colour temperatures (as per Eq. \ref{eq:Tcol}), and a millimetre value of $\beta_\mathrm{mm} = 1.60 \pm 0.06$, for the determination of $\kappa_\mathrm{260}$ used in the mass calculation (Eq. \ref{eq:mass}). In panel f) of Fig. \ref{fig:MTcomparison}, it can be seen that the relative evolutionary trends are unaffected by these changes, with the exception that the value of $\log_{10}(M/M_\odot)$ is systematically shifted downwards by $\sim$0.1, accompanied by a small reduction in temperature of 0.2--0.4 K.

\section{Summary and conclusions} \label{sec:conclusions}

In this paper, we present the Galactic Star Formation with NIKA2 (GASTON) project, a guaranteed-time large programme using the IRAM 30\,m telescope's new millimetre continuum camera, NIKA2. The GASTON large-programme consists of two projects aimed at probing the origin of different regions of the stellar initial mass function: i) a survey of high- to intermediate-mass star-forming clumps within a 2 deg$^2$ region of the inner Galactic plane (GP) centred on $\ell = 23\fdg9, b=0\fdg05$; ii) a search for pre-brown dwarf cores; and a third project: iii) consisting of a study of dust property variations in nearby well-resolved cores. 

We have described the observing strategy for the GASTON GP project, and presented the first results obtained primarily through analysis of the 1.15\,mm photometric maps after $\sim 40$ per cent of the total integration time. A dendrogram extraction technique has allowed us to isolate resolved and unresolved compact sources as well as more extended structures, whilst maintaining a description of the hierarchy of emission complexes. By extracting $^{13}$CO (1--0) spectra for each source using FUGIN \citep{Umemoto2017} and RAMPS \citep{Hogge2018} data in combination with the latest models of the Galactic rotation curve \citep{Reid2019}, we have assigned kinematic distances to each structure and, in concert with Hi-GAL \citep{Molinari2016} data, have been able to calculate their masses.

Of the 2446 dendrogram structures that we have identified, a total of 1467 structures are either compact or unsubstructured (within the constraints of the observations) and 22 per cent of these appear to be a new population of clumps with no counterpart in the \textit{Herschel} Hi-GAL catalogues \citep{Molinari2016, Elia2017}. These new clumps are, on average, less massive, and colder than their 250~\micron-detected counterparts, and represent a population of sources for which the \textit{Herschel} resolution element at wavelengths in excess of 250~\micron\ is not sensitive to. The majority of these sources are candidates for starless clumps, with 80 per cent having no compact 70~\micron counterparts. By the end of the GASTON GP observing campaign, we expect this number to increase significantly.

We have proposed a categorisation that describes the relative evolutionary stage of the clumps in terms of their infrared-bright fraction, $\firb$. The mean temperature of the clumps, along with the fraction of 70~\micron-bright sources, and the ratio bolometric luminosity to mass in each stage increases monotonically with the proposed sequence, supporting the use of $\firb$ as an evolutionary indicator. In mass-temperature space, the mean position of the distribution of clumps follows a path that agrees well with clump-fed scenarios for high-mass star formation, in which high-mass star-forming cores accrete mass from their parsec-scale environment, before finally losing mass at later stages as the dense gas initially associated with the star-forming region is dispersed and accreted.

Upon completion of the GASTON GP field, we will publish the full catalogue of sources, and revisit the measurements we have made in this study.

\section*{Acknowledgements}

We wish to thank the anonymous referee for their helpful comments and suggestions that have improved the quality of this paper. AJR would like to thank his newborn son, Joseph, whose tardy arrival assisted with the completion of this work. AJR and NP would like to thank the STFC for financial support under the consolidated grant number ST/N000706/1. We would like to thank the IRAM staff for their support during the campaigns. The NIKA dilution cryostat has been designed and built at the Institut N\'eel. In particular, we acknowledge the crucial contribution of the Cryogenics Group, and in particular Gregory Garde, Henri Rodenas, Jean Paul Leggeri, Philippe Camus. This work has been partially funded by the Foundation Nanoscience Grenoble and the LabEx FOCUS ANR-11-LABX-0013. This work is supported by the French National Research Agency under the contracts "MKIDS", "NIKA" and ANR-15-CE31-0017 and in the framework of the "Investissements d’avenir” program (ANR-15-IDEX-02). This work has benefited from the support of the European Research Council Advanced Grant ORISTARS under the European Union's Seventh Framework Programme (Grant Agreement no. 291294). F.R. acknowledges financial supports provided by NASA through SAO Award Number SV2-82023 issued by the Chandra X-Ray Observatory Center, which is operated by the Smithsonian Astrophysical Observatory for and on behalf of NASA under contract NAS8-03060. A.B. acknowledges the support of the European Union's Horizon 2020 research and innovation program under the Marie Sk\l{}odowska-Curie Grant agreement No. 843008 (MUSICA). This research made use of the {\sc python} packages {\sc astropy}\footnote{\url{https://astropy.org/}}, \citep{TheAstropyCollaboration2013, TheAstropyCollaboration2018}, {\sc astrodendro}\footnote{\url{http://dendrograms.org/}}, {\sc ipython}\footnote{\url{https://ipython.org/}} \citep{Perez2007}, {\sc numpy}\footnote{\url{https://numpy.org/}}, {\sc scipy}\footnote{\url{https://scipy.org/}} \citep{Virtanen2020}, {\sc matplotlib}\footnote{\url{https://matplotlib.org/}} \citep{Hunter2007} and {\sc multicolorfits}\footnote{\url{https://github.com/pjcigan/multicolorfits}}. This research has also made use of NASA's Astrophysics Data System Bibliographic Services, {\sc topcat}\footnote{\url{http://www.star.bris.ac.uk/~mbt/topcat/}} \citep{Taylor2005}, and {\sc saoimageds9}\footnote{\url{http://ds9.si.edu/}} \citep{Joye2003}.

\section*{Data Availability}

In keeping with the IRAM Large Program policy\footnote{\url{https://www.iram-institute.org/EN/content-page-243-7-158-243-0-0.html}}, the GASTON data -- including raw and processed products -- will be made publicly available 18 months after the last scheduling semester in which it is observed.




\bibliographystyle{mnras}
\bibliography{GASTON_First_Results} 

\begin{thebibliography}{}
\makeatletter
\relax
\def\mn@urlcharsother{\let\do\@makeother \do\$\do\&\do\#\do\^\do\_\do\%\do\~}
\def\mn@doi{\begingroup\mn@urlcharsother \@ifnextchar [ {\mn@doi@}
  {\mn@doi@[]}}
\def\mn@doi@[#1]#2{\def\@tempa{#1}\ifx\@tempa\@empty \href
  {http://dx.doi.org/#2} {doi:#2}\else \href {http://dx.doi.org/#2} {#1}\fi
  \endgroup}
\def\mn@eprint#1#2{\mn@eprint@#1:#2::\@nil}
\def\mn@eprint@arXiv#1{\href {http://arxiv.org/abs/#1} {{\tt arXiv:#1}}}
\def\mn@eprint@dblp#1{\href {http://dblp.uni-trier.de/rec/bibtex/#1.xml}
  {dblp:#1}}
\def\mn@eprint@#1:#2:#3:#4\@nil{\def\@tempa {#1}\def\@tempb {#2}\def\@tempc
  {#3}\ifx \@tempc \@empty \let \@tempc \@tempb \let \@tempb \@tempa \fi \ifx
  \@tempb \@empty \def\@tempb {arXiv}\fi \@ifundefined
  {mn@eprint@\@tempb}{\@tempb:\@tempc}{\expandafter \expandafter \csname
  mn@eprint@\@tempb\endcsname \expandafter{\@tempc}}}

\bibitem[\protect\citeauthoryear{Adam et~al.,}{Adam et~al.}{2018}]{Adam2018}
Adam R.,  et~al., 2018, \mn@doi [A\&A] {10.1051/0004-6361/201731503}, 609, A115

\bibitem[\protect\citeauthoryear{Aguirre et~al.,}{Aguirre
  et~al.}{2011}]{Aguirre2011}
Aguirre J.~E.,  et~al., 2011, \mn@doi [ApJS] {10.1088/0067-0049/192/1/4}, 192,
  4

\bibitem[\protect\citeauthoryear{Andr{\'e} et~al.,}{Andr{\'e}
  et~al.}{2010}]{Andre2010}
Andr{\'e} P.,  et~al., 2010, \mn@doi [A\&A] {10.1051/0004-6361/201014666}, 518,
  L102

\bibitem[\protect\citeauthoryear{Andr{\'e}, Di~Francesco, {Ward-Thompson},
  Inutsuka, Pudritz  \& Pineda}{Andr{\'e} et~al.}{2014}]{Andre2014}
Andr{\'e} P.,  Di~Francesco J.,  {Ward-Thompson} D.,  Inutsuka S.-I.,  Pudritz
  R.~E.,   Pineda J.~E.,  2014, \mn@doi [Protostars and Planets VI]
  {10.2458/azu_uapress_9780816531240-ch002}, pp 27--51

\bibitem[\protect\citeauthoryear{Andr{\'e}, Arzoumanian, K{\"o}nyves, Shimajiri
   \& Palmeirim}{Andr{\'e} et~al.}{2019}]{Andre2019}
Andr{\'e} P.,  Arzoumanian D.,  K{\"o}nyves V.,  Shimajiri Y.,   Palmeirim P.,
  2019, \mn@doi [A\&A] {10.1051/0004-6361/201935915}, 629, L4

\bibitem[\protect\citeauthoryear{Aniano, Draine, Gordon  \& Sandstrom}{Aniano
  et~al.}{2011}]{Aniano2011}
Aniano G.,  Draine B.~T.,  Gordon K.~D.,   Sandstrom K.,  2011, \mn@doi
  [Publications of the Astronomical Society of the Pacific] {10.1086/662219},
  123, 1218

\bibitem[\protect\citeauthoryear{Battersby et~al.,}{Battersby
  et~al.}{2011}]{Battersby2011}
Battersby C.,  et~al., 2011, \mn@doi [A\&A] {10.1051/0004-6361/201116559}, 535,
  A128

\bibitem[\protect\citeauthoryear{Battersby, Bally  \& Svoboda}{Battersby
  et~al.}{2017}]{Battersby2017}
Battersby C.,  Bally J.,   Svoboda B.,  2017, \mn@doi [The Astrophysical
  Journal] {10.3847/1538-4357/835/2/263}, 835, 263

\bibitem[\protect\citeauthoryear{Beckwith, Sargent, Chini  \& Guesten}{Beckwith
  et~al.}{1990}]{Beckwith1990}
Beckwith S. V.~W.,  Sargent A.~I.,  Chini R.~S.,   Guesten R.,  1990, \mn@doi
  [The Astronomical Journal] {10.1086/115385}, 99, 924

\bibitem[\protect\citeauthoryear{Benjamin et~al.,}{Benjamin
  et~al.}{2003}]{Benjamin2003}
Benjamin R.~A.,  et~al., 2003, \mn@doi [PUBL ASTRON SOC PAC] {10.1086/376696},
  115, 953

\bibitem[\protect\citeauthoryear{Bonnell \& Bate}{Bonnell \&
  Bate}{2006}]{Bonnell2006}
Bonnell I.~A.,  Bate M.~R.,  2006, \mn@doi [Monthly Notices of the Royal
  Astronomical Society] {10.1111/j.1365-2966.2006.10495.x}, 370, 488

\bibitem[\protect\citeauthoryear{Bontemps, Motte, Csengeri  \&
  Schneider}{Bontemps et~al.}{2010}]{Bontemps2010}
Bontemps S.,  Motte F.,  Csengeri T.,   Schneider N.,  2010, \mn@doi [A\&A]
  {10.1051/0004-6361/200913286}, 524, A18

\bibitem[\protect\citeauthoryear{Bourrion et~al.,}{Bourrion
  et~al.}{2016}]{Bourrion2016}
Bourrion O.,  et~al., 2016, \mn@doi [Journal of Instrumentation]
  {10.1088/1748-0221/11/11/P11001}, 11, P11001

\bibitem[\protect\citeauthoryear{Calvo et~al.,}{Calvo et~al.}{2016}]{Calvo2016}
Calvo M.,  et~al., 2016, \mn@doi [Journal of Low Temperature Physics]
  {10.1007/s10909-016-1582-0}, 184, 816

\bibitem[\protect\citeauthoryear{Churchwell et~al.,}{Churchwell
  et~al.}{2009}]{Churchwell2009}
Churchwell E.,  et~al., 2009, \mn@doi [Publications of the Astronomical Society
  of the Pacific] {10.1086/597811}, 121, 213

\bibitem[\protect\citeauthoryear{Clarke, Whitworth, {Duarte-Cabral}  \&
  Hubber}{Clarke et~al.}{2017}]{Clarke2017}
Clarke S.~D.,  Whitworth A.~P.,  {Duarte-Cabral} A.,   Hubber D.~A.,  2017,
  \mn@doi [Monthly Notices of the Royal Astronomical Society]
  {10.1093/mnras/stx637}, 468, 2489

\bibitem[\protect\citeauthoryear{Dame, Hartmann  \& Thaddeus}{Dame
  et~al.}{2001}]{Dame2001}
Dame T.~M.,  Hartmann D.,   Thaddeus P.,  2001, \mn@doi [The Astrophysical
  Journal] {10.1086/318388}, 547, 792

\bibitem[\protect\citeauthoryear{Dempsey, Thomas  \& Currie}{Dempsey
  et~al.}{2013}]{Dempsey2013}
Dempsey J.~T.,  Thomas H.~S.,   Currie M.~J.,  2013, \mn@doi [ApJS]
  {10.1088/0067-0049/209/1/8}, 209, 8

\bibitem[\protect\citeauthoryear{Drabek et~al.,}{Drabek
  et~al.}{2012}]{Drabek2012}
Drabek E.,  et~al., 2012, \mn@doi [Monthly Notices of the Royal Astronomical
  Society] {10.1111/j.1365-2966.2012.21140.x}, 426, 23

\bibitem[\protect\citeauthoryear{Dunham, Crapsi, Evans~II, Bourke, Huard, Myers
   \& Kauffmann}{Dunham et~al.}{2008}]{Dunham2008}
Dunham M.~M.,  Crapsi A.,  Evans~II N.~J.,  Bourke T.~L.,  Huard T.~L.,  Myers
  P.~C.,   Kauffmann J.,  2008, \mn@doi [Astrophysical Journal Supplement
  Series, The] {10.1086/591085}, 179, 249

\bibitem[\protect\citeauthoryear{Eden et~al.,}{Eden et~al.}{2020}]{Eden2020}
Eden D.~J.,  et~al., 2020, \mn@doi [Monthly Notices of the Royal Astronomical
  Society] {10.1093/mnras/staa2734}, 498, 5936

\bibitem[\protect\citeauthoryear{Elia et~al.,}{Elia et~al.}{2017}]{Elia2017}
Elia D.,  et~al., 2017, \mn@doi [MNRAS] {10.1093/mnras/stx1357}, 471, 100

\bibitem[\protect\citeauthoryear{{Ellsworth-Bowers} et~al.,}{{Ellsworth-Bowers}
  et~al.}{2013}]{Ellsworth-Bowers2013}
{Ellsworth-Bowers} T.~P.,  et~al., 2013, \mn@doi [ApJ]
  {10.1088/0004-637X/770/1/39}, 770, 39

\bibitem[\protect\citeauthoryear{{Ellsworth-Bowers}, Rosolowsky, Glenn,
  Ginsburg, Evans~II, Battersby, Shirley  \& Svoboda}{{Ellsworth-Bowers}
  et~al.}{2015}]{Ellsworth-Bowers2015}
{Ellsworth-Bowers} T.~P.,  Rosolowsky E.,  Glenn J.,  Ginsburg A.,  Evans~II
  N.~J.,  Battersby C.,  Shirley Y.~L.,   Svoboda B.,  2015, \mn@doi [ApJ]
  {10.1088/0004-637X/799/1/29}, 799, 29

\bibitem[\protect\citeauthoryear{Frayer et~al.,}{Frayer
  et~al.}{2020}]{Frayer2020}
Frayer D.,  et~al., 2020, in Bulletin of the {{American Astronomical Society}}.
  p. 272.17

\bibitem[\protect\citeauthoryear{Ginsburg et~al.,}{Ginsburg
  et~al.}{2013}]{Ginsburg2013}
Ginsburg A.,  et~al., 2013, \mn@doi [ApJS] {10.1088/0067-0049/208/2/14}, 208,
  14

\bibitem[\protect\citeauthoryear{Griffin et~al.,}{Griffin
  et~al.}{2010}]{Griffin2010}
Griffin M.~J.,  et~al., 2010, \mn@doi [A\&A] {10.1051/0004-6361/201014519},
  518, L3

\bibitem[\protect\citeauthoryear{Heyer et~al.,}{Heyer et~al.}{2018}]{Heyer2018}
Heyer M.,  et~al., 2018, \mn@doi [Monthly Notices of the Royal Astronomical
  Society] {10.1093/mnras/stx2484}, 473, 2222

\bibitem[\protect\citeauthoryear{Hogge et~al.,}{Hogge et~al.}{2018}]{Hogge2018}
Hogge T.,  et~al., 2018, \mn@doi [ApJS] {10.3847/1538-4365/aacf94}, 237, 27

\bibitem[\protect\citeauthoryear{Hunter}{Hunter}{2007}]{Hunter2007}
Hunter J.~D.,  2007, \mn@doi [Computing in Science and Engineering]
  {10.1109/MCSE.2007.55}, 9, 90

\bibitem[\protect\citeauthoryear{Inutsuka \& Miyama}{Inutsuka \&
  Miyama}{1997}]{Inutsuka1997}
Inutsuka S.-i.,  Miyama S.~M.,  1997, \mn@doi [ApJ] {10.1086/303982}, 480, 681

\bibitem[\protect\citeauthoryear{Jackson et~al.,}{Jackson
  et~al.}{2006}]{Jackson2006}
Jackson J.~M.,  et~al., 2006, \mn@doi [ASTROPHYS J SUPPL S] {10.1086/500091},
  163, 145

\bibitem[\protect\citeauthoryear{Jackson et~al.,}{Jackson
  et~al.}{2019}]{Jackson2019}
Jackson J.~M.,  et~al., 2019, \mn@doi [The Astrophysical Journal]
  {10.3847/1538-4357/aaef84}, 870, 5

\bibitem[\protect\citeauthoryear{Joye \& Mandel}{Joye \&
  Mandel}{2003}]{Joye2003}
Joye W.~A.,  Mandel E.,  2003, Astronomical Data Analysis Software and Systems
  XII, 295, 489

\bibitem[\protect\citeauthoryear{Kirk, Myers, Bourke, Gutermuth, Hedden  \&
  Wilson}{Kirk et~al.}{2013}]{Kirk2013}
Kirk H.,  Myers P.~C.,  Bourke T.~L.,  Gutermuth R.~A.,  Hedden A.,   Wilson
  G.~W.,  2013, The Astrophysical Journal, p.~14

\bibitem[\protect\citeauthoryear{Klaassen et~al.,}{Klaassen
  et~al.}{2019}]{Klaassen2019}
Klaassen P.,  et~al., 2019, Astro2020: Decadal Survey on Astronomy and
  Astrophysics, APC white papers, no. 58; Bulletin of the American Astronomical
  Society, 51, 58

\bibitem[\protect\citeauthoryear{K{\"o}nyves et~al.,}{K{\"o}nyves
  et~al.}{2015}]{Konyves2015}
K{\"o}nyves V.,  et~al., 2015, \mn@doi [A\&A] {10.1051/0004-6361/201525861},
  584, A91

\bibitem[\protect\citeauthoryear{K{\"o}nyves et~al.,}{K{\"o}nyves
  et~al.}{2020}]{Konyves2020}
K{\"o}nyves V.,  et~al., 2020, \mn@doi [A\&A] {10.1051/0004-6361/201834753},
  635, A34

\bibitem[\protect\citeauthoryear{Lee, Hennebelle  \& Chabrier}{Lee
  et~al.}{2017}]{Lee2017}
Lee Y.-N.,  Hennebelle P.,   Chabrier G.,  2017, \mn@doi [ApJ]
  {10.3847/1538-4357/aa898f}, 847, 114

\bibitem[\protect\citeauthoryear{Liu et~al.,}{Liu et~al.}{2016}]{Liu2016}
Liu T.,  et~al., 2016, \mn@doi [ApJ] {10.3847/0004-637X/829/2/59}, 829, 59

\bibitem[\protect\citeauthoryear{Lu et~al.,}{Lu et~al.}{2018}]{Lu2018}
Lu X.,  et~al., 2018, \mn@doi [ApJ] {10.3847/1538-4357/aaad11}, 855, 9

\bibitem[\protect\citeauthoryear{Luhman}{Luhman}{2012}]{Luhman2012}
Luhman K.~L.,  2012, \mn@doi [Annu. Rev. Astron. Astrophys.]
  {10.1146/annurev-astro-081811-125528}, 50, 65

\bibitem[\protect\citeauthoryear{Marsh et~al.,}{Marsh et~al.}{2016}]{Marsh2016}
Marsh K.~A.,  et~al., 2016, \mn@doi [Mon. Not. R. Astron. Soc.]
  {10.1093/mnras/stw301}, 459, 342

\bibitem[\protect\citeauthoryear{McKee \& Tan}{McKee \& Tan}{2003}]{McKee2003}
McKee C.~F.,  Tan J.~C.,  2003, \mn@doi [ApJ] {10.1086/346149}, 585, 850

\bibitem[\protect\citeauthoryear{Molinari et~al.,}{Molinari
  et~al.}{2010}]{Molinari2010}
Molinari S.,  et~al., 2010, \mn@doi [A\&A] {10.1051/0004-6361/201014659}, 518,
  L100

\bibitem[\protect\citeauthoryear{Molinari et~al.,}{Molinari
  et~al.}{2016a}]{Molinari2016}
Molinari S.,  et~al., 2016a, \mn@doi [A\&A] {10.1051/0004-6361/201526380}, 591,
  A149

\bibitem[\protect\citeauthoryear{Molinari, Merello, Elia, Cesaroni, Testi  \&
  Robitaille}{Molinari et~al.}{2016b}]{Molinari2016a}
Molinari S.,  Merello M.,  Elia D.,  Cesaroni R.,  Testi L.,   Robitaille T.,
  2016b, \mn@doi [ApJ] {10.3847/2041-8205/826/1/L8}, 826, L8

\bibitem[\protect\citeauthoryear{Moore et~al.,}{Moore et~al.}{2015}]{Moore2015}
Moore T. J.~T.,  et~al., 2015, \mn@doi [Mon. Not. R. Astron. Soc.]
  {10.1093/mnras/stv1833}, 453, 4265

\bibitem[\protect\citeauthoryear{Motte, Bontemps  \& Louvet}{Motte
  et~al.}{2018}]{Motte2018a}
Motte F.,  Bontemps S.,   Louvet F.,  2018, \mn@doi [Annu. Rev. Astron.
  Astrophys.] {10.1146/annurev-astro-091916-055235}, 56, 41

\bibitem[\protect\citeauthoryear{Myers}{Myers}{2009}]{Myers2009}
Myers P.~C.,  2009, \mn@doi [ApJ] {10.1088/0004-637X/700/2/1609}, 700, 1609

\bibitem[\protect\citeauthoryear{Pe{\~n}aloza, Clark, Glover, Shetty  \&
  Klessen}{Pe{\~n}aloza et~al.}{2017}]{Penaloza2017}
Pe{\~n}aloza C.~H.,  Clark P.~C.,  Glover S. C.~O.,  Shetty R.,   Klessen
  R.~S.,  2017, \mn@doi [Mon. Not. R. Astron. Soc.] {10.1093/mnras/stw2892},
  465, 2277

\bibitem[\protect\citeauthoryear{Peretto \& Fuller}{Peretto \&
  Fuller}{2009}]{Peretto2009}
Peretto N.,  Fuller G.~A.,  2009, \mn@doi [A\&A] {10.1051/0004-6361/200912127},
  505, 405

\bibitem[\protect\citeauthoryear{Peretto et~al.,}{Peretto
  et~al.}{2013}]{Peretto2013}
Peretto N.,  et~al., 2013, \mn@doi [A\&A] {10.1051/0004-6361/201321318}, 555,
  A112

\bibitem[\protect\citeauthoryear{Peretto et~al.,}{Peretto
  et~al.}{2014}]{Peretto2014}
Peretto N.,  et~al., 2014, \mn@doi [A\&A] {10.1051/0004-6361/201322172}, 561,
  A83

\bibitem[\protect\citeauthoryear{Peretto, Lenfestey, Fuller, Traficante,
  Molinari, Thompson  \& {Ward-Thompson}}{Peretto et~al.}{2016}]{Peretto2016}
Peretto N.,  Lenfestey C.,  Fuller G.~A.,  Traficante A.,  Molinari S.,
  Thompson M.~A.,   {Ward-Thompson} D.,  2016, \mn@doi [A\&A]
  {10.1051/0004-6361/201527064}, 590, A72

\bibitem[\protect\citeauthoryear{Peretto et~al.,}{Peretto
  et~al.}{2020}]{Peretto2020}
Peretto N.,  et~al., 2020, \mn@doi [Monthly Notices of the Royal Astronomical
  Society] {10.1093/mnras/staa1656}, 496, 3482

\bibitem[\protect\citeauthoryear{Perez \& Granger}{Perez \&
  Granger}{2007}]{Perez2007}
Perez F.,  Granger B.~E.,  2007, \mn@doi [Computing in Science Engineering]
  {10.1109/MCSE.2007.53}, 9, 21

\bibitem[\protect\citeauthoryear{Perotto et~al.,}{Perotto
  et~al.}{2020}]{Perotto2020}
Perotto L.,  et~al., 2020, \mn@doi [A\&A] {10.1051/0004-6361/201936220}, 637,
  A71

\bibitem[\protect\citeauthoryear{{Planck Collaboration}}{{Planck
  Collaboration}}{2011}]{PlanckCollaboration2011}
{Planck Collaboration} 2011, \mn@doi [A\&A] {10.1051/0004-6361/201116483}, 536,
  A25

\bibitem[\protect\citeauthoryear{{Planck Collaboration:} et~al.,}{{Planck
  Collaboration:} et~al.}{2014}]{PlanckCollaboration:2014}
{Planck Collaboration:} et~al., 2014, \mn@doi [A\&A]
  {10.1051/0004-6361/201322367}, 564, A45

\bibitem[\protect\citeauthoryear{Poglitsch et~al.,}{Poglitsch
  et~al.}{2010}]{Poglitsch2010}
Poglitsch A.,  et~al., 2010, \mn@doi [A\&A] {10.1051/0004-6361/201014535}, 518,
  L2

\bibitem[\protect\citeauthoryear{Ragan et~al.,}{Ragan et~al.}{2012}]{Ragan2012}
Ragan S.,  et~al., 2012, \mn@doi [A\&A] {10.1051/0004-6361/201219232}, 547, A49

\bibitem[\protect\citeauthoryear{Reid, Dame, Menten  \& Brunthaler}{Reid
  et~al.}{2016}]{Reid2016}
Reid M.~J.,  Dame T.~M.,  Menten K.~M.,   Brunthaler A.,  2016, \mn@doi [ApJ]
  {10.3847/0004-637X/823/2/77}, 823, 77

\bibitem[\protect\citeauthoryear{Reid et~al.,}{Reid et~al.}{2019}]{Reid2019}
Reid M.~J.,  et~al., 2019, \mn@doi [ApJ] {10.3847/1538-4357/ab4a11}, 885, 131

\bibitem[\protect\citeauthoryear{Rigby et~al.,}{Rigby et~al.}{2018}]{Rigby2018}
Rigby A.~J.,  et~al., 2018, \mn@doi [A\&A] {10.1051/0004-6361/201732258}, 615,
  A18

\bibitem[\protect\citeauthoryear{{Roman-Duval}, Jackson, Heyer, Rathborne  \&
  Simon}{{Roman-Duval} et~al.}{2010}]{Roman-Duval2010}
{Roman-Duval} J.,  Jackson J.~M.,  Heyer M.,  Rathborne J.,   Simon R.,  2010,
  \mn@doi [ApJ] {10.1088/0004-637X/723/1/492}, 723, 492

\bibitem[\protect\citeauthoryear{Rosolowsky, Pineda, Kauffmann  \&
  Goodman}{Rosolowsky et~al.}{2008}]{Rosolowsky2008}
Rosolowsky E.~W.,  Pineda J.~E.,  Kauffmann J.,   Goodman A.~A.,  2008, \mn@doi
  [ApJ] {10.1086/587685}, 679, 1338

\bibitem[\protect\citeauthoryear{Ruppin et~al.,}{Ruppin
  et~al.}{2018}]{Ruppin2018}
Ruppin F.,  et~al., 2018, \mn@doi [A\&A] {10.1051/0004-6361/201732558}, 615,
  A112

\bibitem[\protect\citeauthoryear{Sanhueza, Jackson, Zhang, Guzm{\'a}n, Lu,
  Stephens, Wang  \& Tatematsu}{Sanhueza et~al.}{2017}]{Sanhueza2017}
Sanhueza P.,  Jackson J.~M.,  Zhang Q.,  Guzm{\'a}n A.~E.,  Lu X.,  Stephens
  I.~W.,  Wang K.,   Tatematsu K.,  2017, \mn@doi [ApJ]
  {10.3847/1538-4357/aa6ff8}, 841, 97

\bibitem[\protect\citeauthoryear{Schneider et~al.,}{Schneider
  et~al.}{2012}]{Schneider2012}
Schneider N.,  et~al., 2012, \mn@doi [Astronomy and Astrophysics]
  {10.1051/0004-6361/201118566}, 540, L11

\bibitem[\protect\citeauthoryear{Schuller et~al.,}{Schuller
  et~al.}{2009}]{Schuller2009}
Schuller F.,  et~al., 2009, \mn@doi [Astronomy and Astrophysics]
  {10.1051/0004-6361/200811568}, 504, 415

\bibitem[\protect\citeauthoryear{Stanke et~al.,}{Stanke
  et~al.}{2019}]{Stanke2019}
Stanke T.,  et~al., 2019, Astro2020: Decadal Survey on Astronomy and
  Astrophysics, science white papers, no. 542; Bulletin of the American
  Astronomical Society, 51, 542

\bibitem[\protect\citeauthoryear{Svoboda et~al.,}{Svoboda
  et~al.}{2016}]{Svoboda2016}
Svoboda B.~E.,  et~al., 2016, \mn@doi [ApJ] {10.3847/0004-637X/822/2/59}, 822,
  59

\bibitem[\protect\citeauthoryear{Tan, Beltr{\'a}n, Caselli, Fontani, Fuente,
  Krumholz, McKee  \& Stolte}{Tan et~al.}{2014}]{Tan2014}
Tan J.~C.,  Beltr{\'a}n M.~T.,  Caselli P.,  Fontani F.,  Fuente A.,  Krumholz
  M.~R.,  McKee C.~F.,   Stolte A.,  2014, \mn@doi [Protostars and Planets VI]
  {10.2458/azu_uapress_9780816531240-ch007}, pp 149--172

\bibitem[\protect\citeauthoryear{Taylor}{Taylor}{2005}]{Taylor2005}
Taylor M.~B.,  2005, in Astronomical {{Data Analysis Software}} and {{Systems
  XIV}}. p.~29

\bibitem[\protect\citeauthoryear{Terebey, Chandler  \& Andre}{Terebey
  et~al.}{1993}]{Terebey1993}
Terebey S.,  Chandler C.~J.,   Andre P.,  1993, \mn@doi [The Astrophysical
  Journal] {10.1086/173121}, 414, 759

\bibitem[\protect\citeauthoryear{{The Astropy Collaboration} et~al.,}{{The
  Astropy Collaboration} et~al.}{2013}]{TheAstropyCollaboration2013}
{The Astropy Collaboration} et~al., 2013, \mn@doi [A\&A]
  {10.1051/0004-6361/201322068}, 558, A33

\bibitem[\protect\citeauthoryear{{The Astropy Collaboration} et~al.,}{{The
  Astropy Collaboration} et~al.}{2018}]{TheAstropyCollaboration2018}
{The Astropy Collaboration} et~al., 2018, \mn@doi [AJ]
  {10.3847/1538-3881/aabc4f}, 156, 123

\bibitem[\protect\citeauthoryear{Traficante, Fuller, Billot, {Duarte-Cabral},
  Merello, Molinari, Peretto  \& Schisano}{Traficante
  et~al.}{2017}]{Traficante2017}
Traficante A.,  Fuller G.~A.,  Billot N.,  {Duarte-Cabral} A.,  Merello M.,
  Molinari S.,  Peretto N.,   Schisano E.,  2017, \mn@doi [Monthly Notices of
  the Royal Astronomical Society] {10.1093/mnras/stx1375}, 470, 3882

\bibitem[\protect\citeauthoryear{Umemoto et~al.,}{Umemoto
  et~al.}{2017}]{Umemoto2017}
Umemoto T.,  et~al., 2017, \mn@doi [Publications of the Astronomical Society of
  Japan] {10.1093/pasj/psx061}, 69

\bibitem[\protect\citeauthoryear{Urquhart et~al.,}{Urquhart
  et~al.}{2014}]{Urquhart2014}
Urquhart J.~S.,  et~al., 2014, \mn@doi [A\&A] {10.1051/0004-6361/201424126},
  568, A41

\bibitem[\protect\citeauthoryear{Urquhart et~al.,}{Urquhart
  et~al.}{2018}]{Urquhart2018}
Urquhart J.~S.,  et~al., 2018, \mn@doi [Monthly Notices of the Royal
  Astronomical Society] {10.1093/mnras/stx2258}, 473, 1059

\bibitem[\protect\citeauthoryear{{V{\'a}zquez-Semadeni}, G{\'o}mez, Jappsen,
  {Ballesteros-Paredes}  \& Klessen}{{V{\'a}zquez-Semadeni}
  et~al.}{2009}]{Vazquez-Semadeni2009}
{V{\'a}zquez-Semadeni} E.,  G{\'o}mez G.~C.,  Jappsen A.-K.,
  {Ballesteros-Paredes} J.,   Klessen R.~S.,  2009, \mn@doi [ApJ]
  {10.1088/0004-637X/707/2/1023}, 707, 1023

\bibitem[\protect\citeauthoryear{{V{\'a}zquez-Semadeni}, Palau,
  {Ballesteros-Paredes}, G{\'o}mez  \&
  {Zamora-Avil{\'e}s}}{{V{\'a}zquez-Semadeni}
  et~al.}{2019}]{Vazquez-Semadeni2019}
{V{\'a}zquez-Semadeni} E.,  Palau A.,  {Ballesteros-Paredes} J.,  G{\'o}mez
  G.~C.,   {Zamora-Avil{\'e}s} M.,  2019, \mn@doi [Monthly Notices of the Royal
  Astronomical Society] {10.1093/mnras/stz2736}, 490, 3061

\bibitem[\protect\citeauthoryear{Virtanen et~al.,}{Virtanen
  et~al.}{2020}]{Virtanen2020}
Virtanen P.,  et~al., 2020, \mn@doi [Nature Methods]
  {10.1038/s41592-019-0686-2}, 17, 261

\bibitem[\protect\citeauthoryear{Wang, Li, Abel  \& Nakamura}{Wang
  et~al.}{2010}]{Wang2010}
Wang P.,  Li Z.-Y.,  Abel T.,   Nakamura F.,  2010, \mn@doi [ApJ]
  {10.1088/0004-637X/709/1/27}, 709, 27

\bibitem[\protect\citeauthoryear{Whitworth, Bate, Nordlund, Reipurth  \&
  Zinnecker}{Whitworth et~al.}{2007}]{Whitworth2007}
Whitworth A.,  Bate M.~R.,  Nordlund {\AA}.,  Reipurth B.,   Zinnecker H.,
  2007, Protostars and Planets V, pp 459--476

\bibitem[\protect\citeauthoryear{Williams, Peretto, Avison, {Duarte-Cabral}  \&
  Fuller}{Williams et~al.}{2018}]{Williams2018}
Williams G.~M.,  Peretto N.,  Avison A.,  {Duarte-Cabral} A.,   Fuller G.~A.,
  2018, \mn@doi [A\&A] {10.1051/0004-6361/201731587}, 613, A11

\makeatother
\end{thebibliography}




\appendix

\section{Null maps} \label{app:nullmaps}

\begin{figure}
    \centering
    \includegraphics[width=\linewidth]{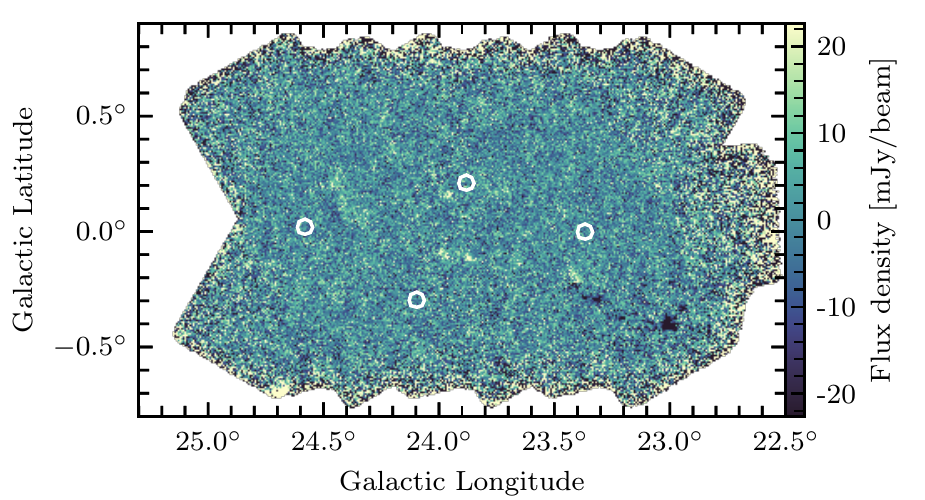}
    \caption{Null map for the GASTON Galactic Plane field at 1.15\,mm. The positions of the four sky apertures are shown as white polygons.}
    \label{fig:nullmaps}
\end{figure}

We display the 1,15\,mm null map, used in Sect. \ref{sec:sourceextraction}, in Fig. \ref{fig:nullmaps}.

\section{Basic source properties} \label{app:catalogue}

\begin{figure*}
    \centering
    \includegraphics[width=\linewidth]{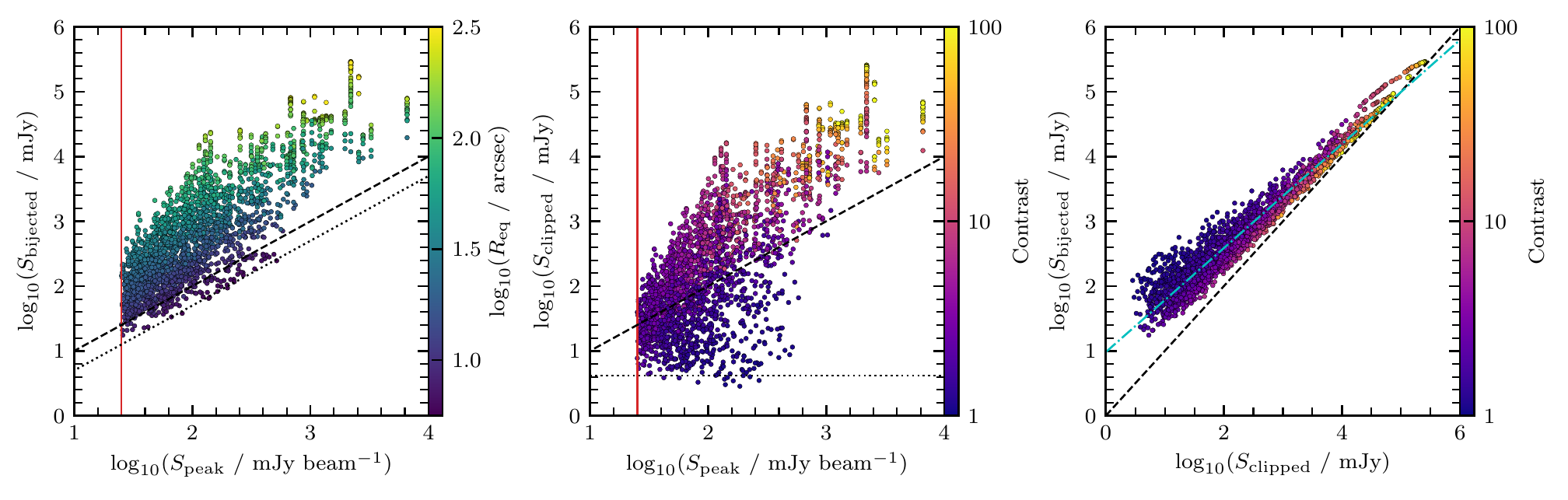}
    \caption{Left panel: peak versus integrated (bijected) flux density for all sources, colour-coded according to sources' equivalent radii. The dashed 1:1 line shows the expected relationship for isolated point sources in all panels. The dotted line shows the lower bound of the data, where the integrated flux density is equal to half of the expected value for a point source. Middle panel: peak versus integrated (clipped) flux density for all sources, colour-coded according to the contrast parameter (see text). The dashed black line again shows the expected relationship for isolated point sources. The dotted line shows the minimum integrated flux density of a point source at the detection threshold. Right panel: clipped versus bijected integrated flux density, coloured by contrast parameter.}
    \label{fig:cataloguefluxes}
\end{figure*}

\begin{figure*}
    \includegraphics[width=\linewidth]{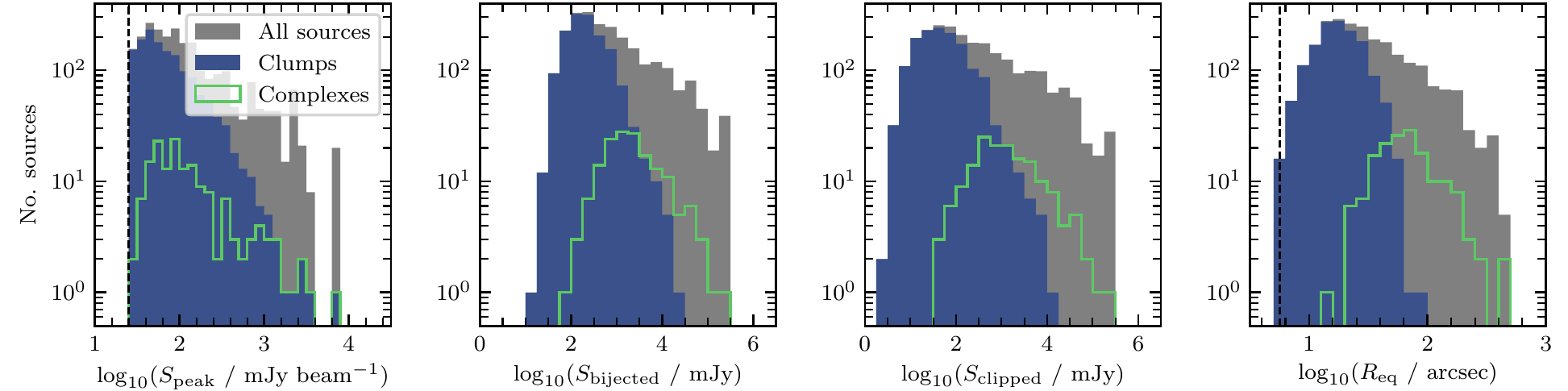}
    \caption{Distributions of various source properties, separated into all sources (grey), clumps (sources containing no discernible substructure), and complexes (sources containing two or more clumps).}
    \label{fig:cataloguehists}
\end{figure*}

In Section \ref{sec:sourceextraction}, we described the dendrogram-based procedure used to identify discrete sources within the 1.15\,mm data, and here we describe their basic properties. We have measured two types of integrated flux density, termed `bijected' and `clipped', as in the scheme of \citet{Rosolowsky2008}. The bijected integrated flux density of a source is calculated from the sum of pixel values within the source. By contrast, the clipped integrated flux density is a background-subtracted measurement, in which an estimate of the local background (as described in Sect. \ref{sec:properties}) is first subtracted from each pixel value before calculating their sum. We further define a contrast parameter for the clipped flux densities, which is the ratio of the peak to the background flux density value, and has a value of unity for faint sources at the detection threshold of our dendrogram extraction. None of the sources in our dendrogram extraction have a contrast value of less than unity although, in principle, point sources located in regions with a bright background could result in values of less than one.

Figure \ref{fig:cataloguefluxes} illustrates the relationship between peak flux density, and the two types of integrated flux density. The relationship between peak and bijected flux densities shows that the majority of sources lie above the line of equality, which isolated point sources at very high S/N would be expected to follow. Sources lying above this line are extended sources, as can be seen by the colour scale indicating the sources' equivalent radii, $R_\mathrm{eq}$ (the radius of a circle with the equivalent area). In the case of the clipped flux densities, sources lying in regions with significant backgrounds may lie below the line of equality, and indeed the most significant departures occur for sources located in the region around $\ell = 23\fdg4$, $b=-0\fdg2$. The minimum expected clipped integrated flux density is 4.1 mJy, for point sources at the detection threshold, which could occur at any value of peak intensity if the local background is sufficiently bright. Some sources lie just below this line, which may occur where sources are either not point-like, or have local noise levels slightly higher than the adopted average rms in the extraction parameters. The relationship between clipped and bijected integrated intensities is also shown, in which it can be seen that sources with the highest contrast, i.e., sources in regions with relatively low backgrounds, lie closer to the 1:1 relation. A linear least-squares fit in log-space finds an average relationship between the clipped and bijected flux densities of $S_\mathrm{bijected} = 9.6 S_\mathrm{clipped}^{0.8}$.

In Fig. \ref{fig:cataloguehists} we display histograms of the peak and integrated flux densities, and equivalent radii of all sources identified by the dendrogram. The distributions of further subsamples are shown, and we show the population of clumps (sources containing no further observed substructures), and complexes (dendrogram 'trunks' containing more than one clump). Note that in the case of the peak flux densities, the peak flux within a complex is the same value as the peak flux of its brightest constituent clump, and so the complex distribution is a sub-set of the clump distribution. The brightest clumps have peak flux densities of $\sim10$ Jy\,beam$^{-1}$, and the largest complexes have integrated flux densities of $\sim 100$ Jy. Source sizes range from the minimum detectable size of $R_\mathrm{eq} = 5.6$ arcsec to 6.5 arcmin.

\section{Velocity assignments} \label{app:velocities}

\begin{figure*}
    \centering
    \begin{subfigure}{\textwidth}
    \includegraphics[width=\textwidth]{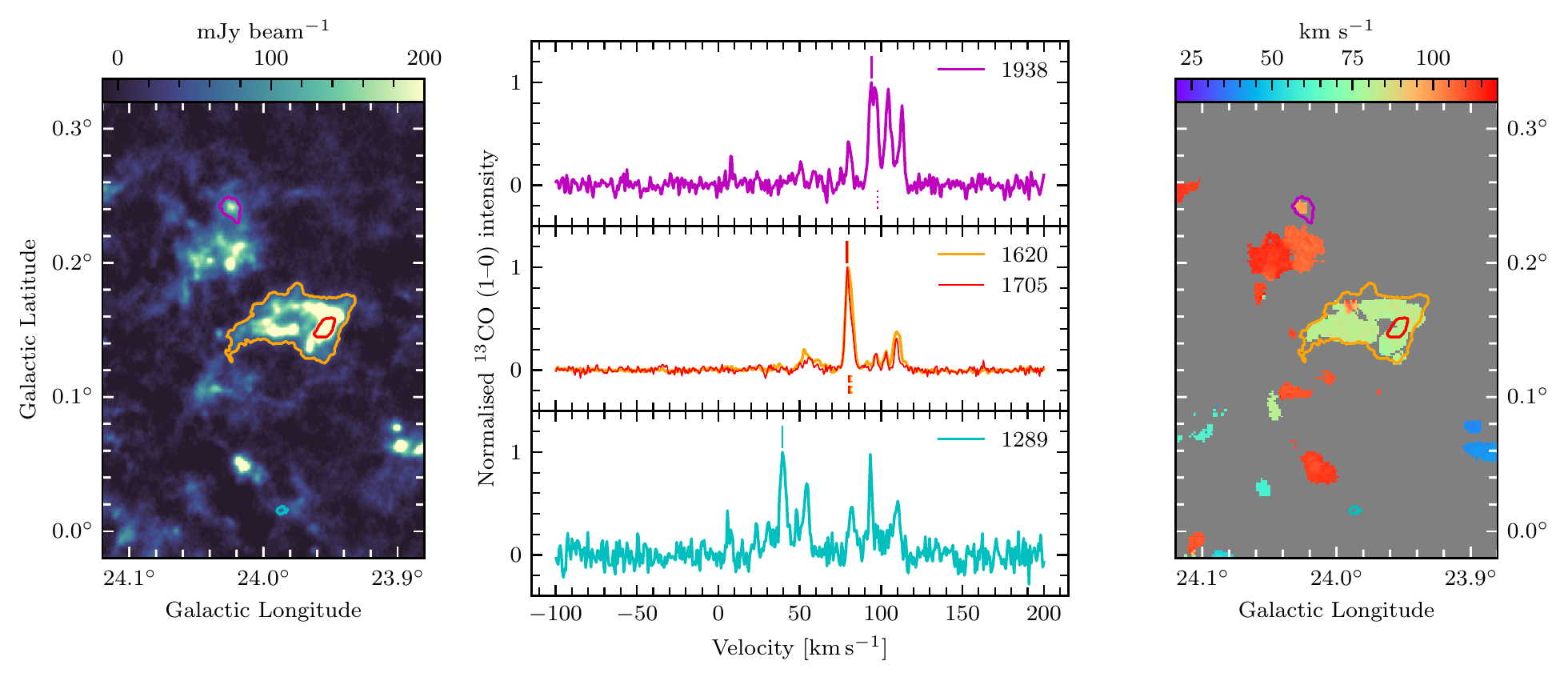}
    \end{subfigure}
    \bigskip
    \begin{subfigure}{\textwidth}
    \centering
    \begin{tabular}{cccccc}
    \hline
    ID & $v_\mathrm{CO}$ & flag & $S/N$ & $v_\mathrm{m1}$ & $v_\mathrm{c1}$ \\
    & (km~s$^{-1}$) &  &  & (km~s$^{-1}$)  & (km~s$^{-1}$) \\
    \hline
    1289 & 39.4$^*$ & 1 & 10.3 & -- & -- \\
    1705 & 79.1 & 3 & 41.0 & 80.3 & 80.2$^*$ \\
    1620 & 79.7 & 3 & 75.7 & 81.7 & 81.1$^*$ \\
    1938 & 94.0 & 2 & 17.1 & 97.8$^*$ & -- \\

    \hline
    \end{tabular}
    \end{subfigure}
    \caption{Example of the velocity determinations of four sources within a sub-set of the data. In the left panel, the 1.15\,mm flux density map is shown (after smoothing to 13 arcsec resolution), with contours overlaid corresponding to the footprints of catalogued sources with ID numbers 1289, 1620, 1705, and 1938. In the central panel, the corresponding $^{13}$CO (1--0) spectra are presented, which were extracted over the footprint of the sources, with the velocity associated with the maximum intensity is assigned to that source indicated by a solid line above the spectra, and any velocity information determined from either the first-fitted velocity component or first moment maps of the RAMPS data are indicated by dot-dashed and dotted lines below the spectra. The right-hand image shows the RAMPS NH$_3$ (1,1) first moment (intensity weighted coordinate) map from \citet{Hogge2018}, with the same source outlines overlaid. In the Table, details of the $^{13}$CO-derived velocity ($v_\mathrm{CO}$), initial velocity flag, the S/N of the $^{13}$CO (1--0) spectrum, and the source velocities derived from the RAMPS NH$_{3}$ (1,1) first moment ($v_\mathrm{m1}$) and first velocity component ($v_\mathrm{c1}$), are given for each source. In each case, an asterisk denotes the adopted velocity, $v_\mathrm{cen}$.}
    \label{fig:spectra}
\end{figure*}

In Sect. \ref{sec:velocities}, we derived an initial velocity assignment for each source based on the channel within a source area-averaged spectrum with the maximum intensity. A level of robustness for each velocity assignment was estimated by determining the standard deviation of the velocities associated with the three brightest channels within each spectrum. In cases where this standard deviation is less than 1~\kms (i.e. the three brightest channels are almost perfectly adjacent), the assignment is considered to be robust and assigned a flag value 3; where the standard deviation is more than 5~\kms, the assignment is considered to be poor and assigned a flag value of 1; cases between those limits are assigned an intermediate robustness flag value of 2. In practice, these flags allow the most robust velocities -- where there is a single dominant peak -- to be isolated from cases with either a low S/N or from spectra in which there are multiple velocity components of similar strength. In this initial velocity assignment, we record 1680 (69 per cent) of sources with a robustness flag of 3, 305 (12 per cent) sources with an intermediate robustness flag, and 461 (19 per cent) with a poor robustness flag.

Following this initial velocity assignment, we make further velocity estimates by using the more appropriate -- but limited in coverage -- data of the NH$_3$ (1,1) inversion transition from the RAMPS survey. The RAMPS pilot study \citep{Hogge2018} presented, as data products, maps of source velocities derived from a line-fitting procedure, as well maps of the first moment, with the latter having a greater extent, but the former being more accurate. For each GASTON source, and for each RAMPS velocity map, we record the mean velocity of pixels that fall within the source area. After excluding measurements where either the mean error (in the case of the fitted first components) or the standard deviation on the mean velocities (in the case of the first moments) are greater than 5~\kms, we record a total of 523 first-component velocities and 1097 first moment velocities. Where available, we replace the $^{13}$CO-derived $v_\mathrm{cen}$ values with their NH$_3$-derived counterparts, preferring the line-fitted measurements over the first moment-derived velocities, and adopt a robustness flag of 3. At this stage, the $v_\mathrm{cen}$ assignments are made up of 1407 (58 per cent) $^{13}$CO-derived velocities, 516 (21 per cent) with NH$_3$ first moment-derived velocities, and the remaining 523 (21 per cent) with velocities derived from the NH$_3$ fitted first velocity components. There are 1913 (78 per cent), 174 (7 per cent) and 359 (15 per cent) of $v_\mathrm{cen}$ assignments with quality flags of 3, 2, and 1, respectively. 888 (85 per cent) of the NH$_3$-derived velocities agree with their initial $^{13}$CO-derived assignment to better than 5~\kms, indicating that the initial assignments are generally reliable.

In Fig. \ref{fig:spectra}, we illustrate the method used in Sect. \ref{sec:velocities} for the initial velocity assignments for each of the sources identified in Sect. \ref{sec:sourceextraction}. Where possible, we also display the velocity derived from the available RAMPS data. At this point, we refer the reader back to the main text of Sect. \ref{sec:velocities} for a description of how the emission structures are used to refine the individual velocity measurements.

\section{Line contamination} \label{app:contamination}

The 1.15 mm NIKA2 band spans a range in frequency from 210--300 GHz (taking these edges as 20 per cent transmission), and as such, the continuum may contain important contributions from the CO (2--1) line at 230.538 GHz, especially for observations of the inner Galactic plane. To estimate the level of this line contamination in the 1.15 mm maps, we made use of CO (1--0) data from the FOREST unbiased Galactic plane imaging survey with the Nobeyama 45 m telescope \citep[FUGIN][]{Umemoto2017}. FUGIN covers the entire GASTON GP region, with an angular resolution of 20 arcsec.

The CO (1--0) data were integrated over the full velocity range of $-50$ to 200~km~s$^{-1}$, and multiplied by a line ratio of $R_{2-1/1-0} = 0.7$ to provide an approximate conversion to CO (2--1) that is appropriate for cold and dense molecular gas \citep{Penaloza2017}. The data were then converted from units of integrated intensity (K~km~s$^{-1}$) to surface brightness units (Jy~beam$^{-1}$) following the procedure outlined in \citet{Drabek2012}, which accounts for the NIKA2 filter response. Finally, we applied a 6.5 arcmin filter to the CO data to approximate the spatial filtering applied by the NIKA2 data reduction pipeline, using the {\sc nebuliser} application. Both the CO (2--1) and NIKA2 1.15~mm surface brightness maps were smoothed to a common resolution of 22 arcsec, and we measured their ratio for pixels that had a 1.15\,mm S/N of more than 10.

The level of CO (2--1) line contamination in the valid pixels ranges from 0 to $\sim$15 per cent, with a median value of 4 per cent. The upper limit of this measurement is most likely to be a worst-case scenario, and would apply in cases where NIKA2 is fully recovering emission up to 6.5\arcmin\ scales. Furthermore, the adopted $R_{2-1/1-0}$ value is the higher value from what is usually measured as a bimodal distribution, with a lower value of 0.3 often found in regions of sub-thermally excited low density CO. We do not, therefore, directly account for CO (2--1) line contamination in our analysis, but incorporate an additional factor of 4 per cent as the associated uncertainty on the flux density.

We note that these values are comparable to those estimated by \citep{Rigby2018} for NIKA, who estimated CO contamination at the 1--3 per cent level. A slightly higher value is found here for a number of reasons: i) the FUGIN data allow us to estimate the contamination at higher angular resolution; ii) The FUGIN data allow us to estimate this quantity in the correct isotopologue of CO whereas the \citet{Rigby2018} study used an approximate conversion from $^{13}$CO, which is less effected by optical depth effects; iii) larger spatial frequencies are able to recovered with NIKA2 than with NIKA, and so more large-scale CO emission survives the pipeline's filtering.

\section{Distance comparison} \label{app:distancecomparison}

\begin{figure}
    \includegraphics[width=\linewidth]{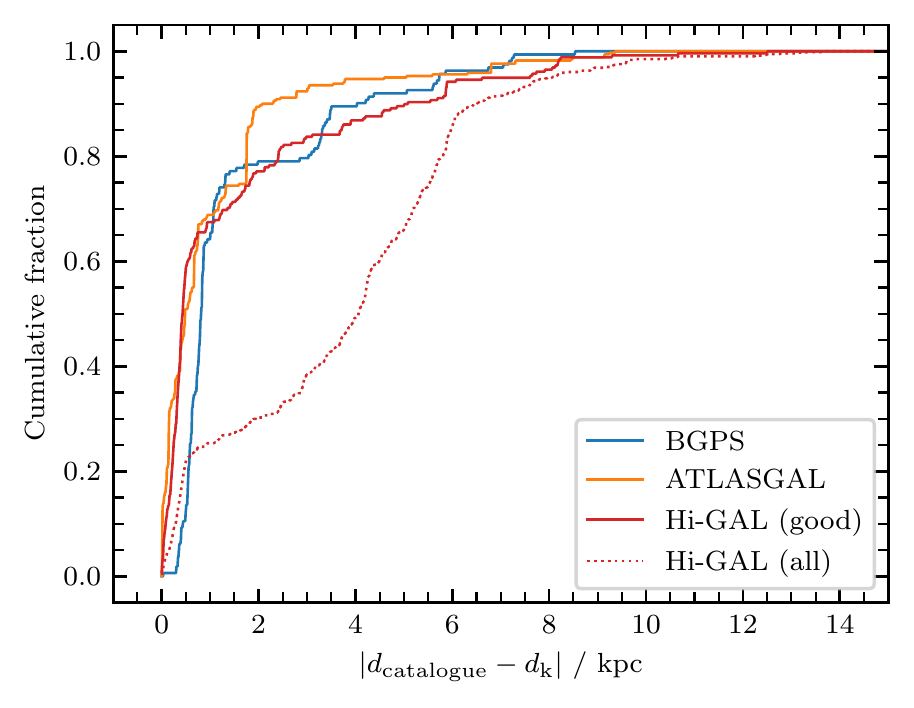}
    \caption{Cumulative distribution showing the difference in distances determined for GASTON clumps and their counterparts from the BGPS (blue), ATLASGAL (orange), and Hi-GAL (red) catalogues.}
    \label{fig:distancecomparison}
\end{figure}

In Fig. \ref{fig:distancecomparison}, we compare the distances determined for GASTON clumps in Sect. \ref{sec:distances} with the distances to their closest positionally-matched counterparts from the BGPS, ATLASGAL, and Hi-GAL surveys, whose catalogues were cross-matched in Sect. \ref{sec:comparison}. We find that 64 and 69 per cent of GASTON clumps have distances that differ by less than 1 kpc when compared to the clump catalogues from the BGPS \citep{Ellsworth-Bowers2015} and ATLASGAL \citep{Urquhart2018}, respectively. While only 25 per cent of distance determinations agree with Hi-GAL clumps from the band-merged catalogue of \citet{Elia2017}, this value rises to 67 per cent once sources without a `good' distance quality flag are removed.

\vspace{0.3cm}

\noindent\textit{$^{1}$School of Physics and Astronomy, Cardiff University, Queen's Buildings, The Parade, Cardiff, CF24 3AA, UK\\
$^{2}$LLR (Laboratoire Leprince-Ringuet), CNRS, École Polytechnique, Institut Polytechnique de Paris, Palaiseau, France\\
$^{3}$AIM, CEA, CNRS, Universit\'e Paris-Saclay, Universit\'e Paris Diderot, Sorbonne Paris Cit\'e, 91191 Gif-sur-Yvette, France\\
$^{4}$Univ. Grenoble Alpes, CNRS, IPAG, 38000 Grenoble, France\\
$^{5}$Institut d'Astrophysique Spatiale (IAS), CNRS and Universit\'e Paris Sud, Orsay, France\\
$^{6}$Institut N\'eel, CNRS and Universit\'e Grenoble Alpes, France\\
$^{7}$Institut de RadioAstronomie Millim\'etrique (IRAM), Grenoble, France\\
$^{8}$Univ. Grenoble Alpes, CNRS, Grenoble INP, LPSC-IN2P3, 53, avenue des Martyrs, 38000 Grenoble, France\\
$^{9}$Rudjer Bo\v{s}kovi\'{c} Institute, Bijeni\v{c}ka cesta 54, 10000 Zagreb, Croatia\\
$^{10}$Dipartimento di Fisica, Sapienza Universit\`a di Roma, Piazzale Aldo Moro 5, I-00185 Roma, Italy\\
$^{11}$Chinese Academy of Sciences South America Center for Astronomy, National Astronomical Observatories, CAS, Beijing 100101, PR China\\
$^{12}$Instituto de Astronom\'ia, Universidad Cat\'olica del Norte, Av. Angamos 0610, Antofagasta 1270709, Chile\\
$^{13}$Centro de Astrobiolog\'ia (CSIC-INTA), Torrej\'on de Ardoz, 28850 Madrid, Spain\\
$^{14}$Instituto de Radioastronom\'ia Milim\'etrica (IRAM), Granada, Spain\\
$^{15}$Aix Marseille Univ, CNRS, CNES, LAM (Laboratoire d'Astrophysique de Marseille), Marseille, France\\
$^{16}$LERMA, Observatoire de Paris, PSL Research University, CNRS, Sorbonne Universit\'es, UPMC Univ. Paris 06, 75014 Paris, France\\
$^{17}$School of Earth and Space Exploration and Department of Physics, Arizona State University, Tempe, AZ 85287, USA\\
$^{18}$Univ. Toulouse, CNRS, IRAP, 9 Av. du Colonel Roche, BP 44346, 31028, Toulouse, France\\
$^{19}$D\'{e}partement de Physique, Ecole Normale Sup\'{e}rieure, 24, rue Lhomond 75005 Paris, France\\
$^{20}$Department of Physics and Astronomy, University of Pennsylvania, 209 South 33rd Street, Philadelphia, PA, 19104, USA\\
$^{21}$Institut d'Astrophysique de Paris, Sorbonne Universit\'{e}, CNRS UMR7095, 75014 Paris, France\\
$^{22}$Kavli Institute for Astrophysics and Space Research, Massachusetts Institute of Technology, Cambridge, MA 02139, USA\\
$^{23}$Astronomisches Rechen-Institut, Zentrum f\:{u}r Astronomie der Universit\:{a}t Heidelberg, M\:{o}nchhofstra\ss e 12-14, D-69120 Heidelberg, Germany
}


\bsp	
\label{lastpage}
\end{document}